# Extreme mixing enables broadband high-temperature electromagnetic absorption in high-entropy perovskites

Ziyu Feng[1], Fangchao Gu[1], Lei Zhuang[1*], Wu Wang[2*], Hulei Yu[1], Yanhui Chu[1*]

[1]School of Materials Science and Engineering, South China University of Technology, Guangzhou, 510641, China.

[2]College of Integrated Circuits and Optoelectronic Chips, Shenzhen Technology University, Shenzhen, 518118, China.

[*]Corresponding author. lzhuang@scut.edu.cn (L.Z); wangwu@sztu.edu.cn (W.W); chuyh@scut.edu.cn (Y.C).

**Abstract**

Developing oxide absorbers with broadband high-temperature electromagnetic absorption performance has long been desirable in the electromagnetic absorption field, yet their absorption performance has remained unsatisfactory. Here, through an extreme mixing strategy, we successfully develop an ultrabroad effective absorption bandwidth (EAB) of 9.3 GHz in high-entropy perovskites up to 600 °C, surpassing the previously reported oxide absorbers. Specifically, we achieve the extreme mixing in high-entropy perovskites by the successful incorporation of up to 13 cation elements, with 8 and 5 cations occupying the A- and B-sites, respectively, using a laser-driven controllable synthesis technique. Such extreme mixing maximally promotes the formation of atomic-scale interfaces to markedly amplify atomic-level interfacial polarization loss for the effective dissipation of electromagnetic wave energy, thereby giving rise to a remarkably broad EAB of 9.3 GHz in high-entropy perovskites up to 600 °C. Owing to the profoundly increased electrical conductivity that causes impedance mismatch at elevated temperatures, however, their EAB deteriorates to 4.6 GHz at 700 °C. This work establishes a new extreme mixing design paradigm for enabling high-entropy oxide absorbers with unprecedented high-temperature EM absorption performance.

## Introduction

As a critical solution to the proliferation of EM pollution and the threat of radar detection, electromagnetic (EM) absorption technology is gaining growing significance in civilian and military applications.[1,2] In recent years, the rapid evolution of the technology has not only expanded the effective frequency range, but also extended its applications from ambient to high-temperature environments—such as radar-stealth aerospace vehicles, as well as high-power electronics, radar stations, and nuclear plants that require EM interference shielding at elevated temperatures.[3-5] This, in turn, calls for electromagnetic wave (EMW) absorbers with broadband high-temperature absorption performance. Ceramic absorbers with chemical inertness, thermal stability, and temperature-insensitive EMW dissipation capability are highly attractive for EMW absorption applicaitons,[6,7] particularly in high-temperature environments where magnetic alloys are limited by Curie temperature and polymers suffer from thermal decomposition.[8,9] Typically, ceramic absorbers are classified into non-oxide and oxide types. Non-oxide absorbers (e.g., carbides[10] and sulfides[11]) commonly exhibit a relatively broad effective absorption bandwidth (EAB), yet they are plagued by oxidation-induced performance deterioration, limiting their applications in oxidizing environments.[12,13] By contrast, oxides stand out as exceptionally promising alternatives for such scenarios, owing to their prominent oxidation inertness. Unfortunately, this advantage is largely offset by their insufficient EMW energy-dissipation pathways, which confine their EAB to below 2 GHz.[14] To overcome this constraint, a variety of structural designs have been explored across multiple length scales—from macroscale metasurfaces for frequency-selective resonant absorption,[15,16] to microscale hollow architectures for multiple scattering,[17] and to nanoscale core–shell heterointerfaces for interfacial polarization.[18] Despite these efforts, the EAB of oxide absorbers rarely exceeds 6 GHz, which still impedes their practical deployment.

A conceptually new pathway has emerged with the high-entropy design strategy, which enables the regulation of atomic-scale configurations.[19,20] Particularly, early investigations have indicated that this strategy holds promise for improving the EMW absorption properties of ceramic absorbers. For example, a (Ti, Zr, Cr Nb, Ta)C high-entropy carbide (HEC) was shown to exhibit enhanced EMW absorption performance relative to its parent components.[21] Similarly, a $(Ti, Nb, Ta)_2FeC$ MAX phase was reported to possess a broadened EAB.[22] However, due to the unresolved critical puzzle regarding the atomic-scale mechanisms of EMW dissipation, progress in expanding the EAB beyond 7 GHz had been hindered. Recently, by using the high-entropy strategy, we achieved a broad EAB of 7.2 GHz in high-entropy diborides (HEBs) and identified that vacancies and atomic-scale interfaces are crucially responsible for the EAB enhancement.[23] A following-up systematic single-variable study revealed that the EM absorption performance is primarily governed by lattice-distortion-driven atomic-scale interfaces, which induce substantial atomic-level interfacial polarization loss to dissipate EMW energy.[24] Consequently, intensifying lattice distortion to maximize the formation of atomic-scale interfaces should be an attractive avenue for further boosting the EMW absorption performance of ceramic absorbers, though it remains highly challenging. Notably, an extreme mixing concept has recently emerged in the alloy community.[25] Leveraging extreme driving force enabled by ultrahigh-temperature synthesis and elevated configurational entropy by increasing the number of constituent elements to offset enthalpy penalty, immiscible elements can be incorporated into one high-entropy alloy lattice, opening a new route to intensify lattice distortion. As such, this strategy holds great potential to maximize atomic-scale interfaces, amplifying interfacial polarization loss and unlocking unprecedented EAB gains in high-entropy oxide absorbers. However, related work is still lacking.

In this work, to implement the extreme mixing strategy, we select $ABO_3$-type perovskite oxides as a platform, taking advantage of their vast compositional tunability across both A- and B-sites, as well as their remarkable cation-size tolerance via octahedral distortion and tilting—which may accommodate the substantial intensification of lattice distortion. As a result, we achieve a broad EAB of 9.3 GHz in high-entropy perovskite oxides (HEPOs), setting a record for oxide-based EMW absorbers. This significant EAB gain predominantly arises from maximizing lattice distortion through the incorporation of up to 13 cation elements in HEPOs (denoted as 13HEPO-AB)—8 and 5 cations at the A- and B-sites, respectively, using a laser-driven controllable synthesis technique. Such extreme mixing markedly promotes the formation of atomic-scale interfaces, thus amplifying atomic-level polarization loss for dissipating EM energy. Notably, the as-developed 13HEPO-AB absorber maintains the broad EAB of 9.3 GHz across room temperature (RT) to 600 °C, underscoring a 100% EAB retention—a performance that surpasses all previously reported oxide absorbers. But at 700 °C, the EAB deteriorates to 4.6 GHz, mainly due to the impedance mismatch caused by thermally increased electrical conductivity ($\sigma$). This work establishes a robust extreme mixing paradigm for developing oxide absorbers capable of operating in high-temperature oxidizing environments.

**Results and discussion**

To implement the extreme mixing strategy in HEPO absorbers, it is imperative to distinguish the individual contributions of the A- and B-sites to the EMW absorption properties. To this end, we designed two sets of HEPO samples with independent compositional engineering at each site, denoted as HEPO-A and HEPO-B (Table S1 and S2). First, the synthesis difficulty of these HEPO samples was evaluated via the size-mismatch factor ($\delta$), according to the Hume-Rothery solid solution rule.[26,27] Fig. 1a shows that $\delta$ rises steadily with increasing number of elements from 4 to 10 at both sites (labeled as 4–10HEPO-A and 4–10HEPO-B, respectively), reflecting the growing synthesis difficulty with expanding compositional complexity. Experimentally, we utilized a laser-driven controllable technique to realize the synthesis.[28] XRD patterns (Fig. 1b, c) along with Rietveld refinement profiles (Fig. S1) suggest that the single-phase 4–10HEPO-A and 4–8HEPO-B samples (space group $Pm\bar{3}m$) have been successfully synthesized. It is noteworthy that the broader compositional flexibility of the A-site compared with the B-site—as evidenced by phase separation in the 9- and 10HEPO-B samples (Fig. S2)—mainly originates from the larger free volume at the A-site available for cations. By contrast, the B-site is confined to much smaller octahedral interstices, which thereby exclude cations with larger ionic radii.[29,30]

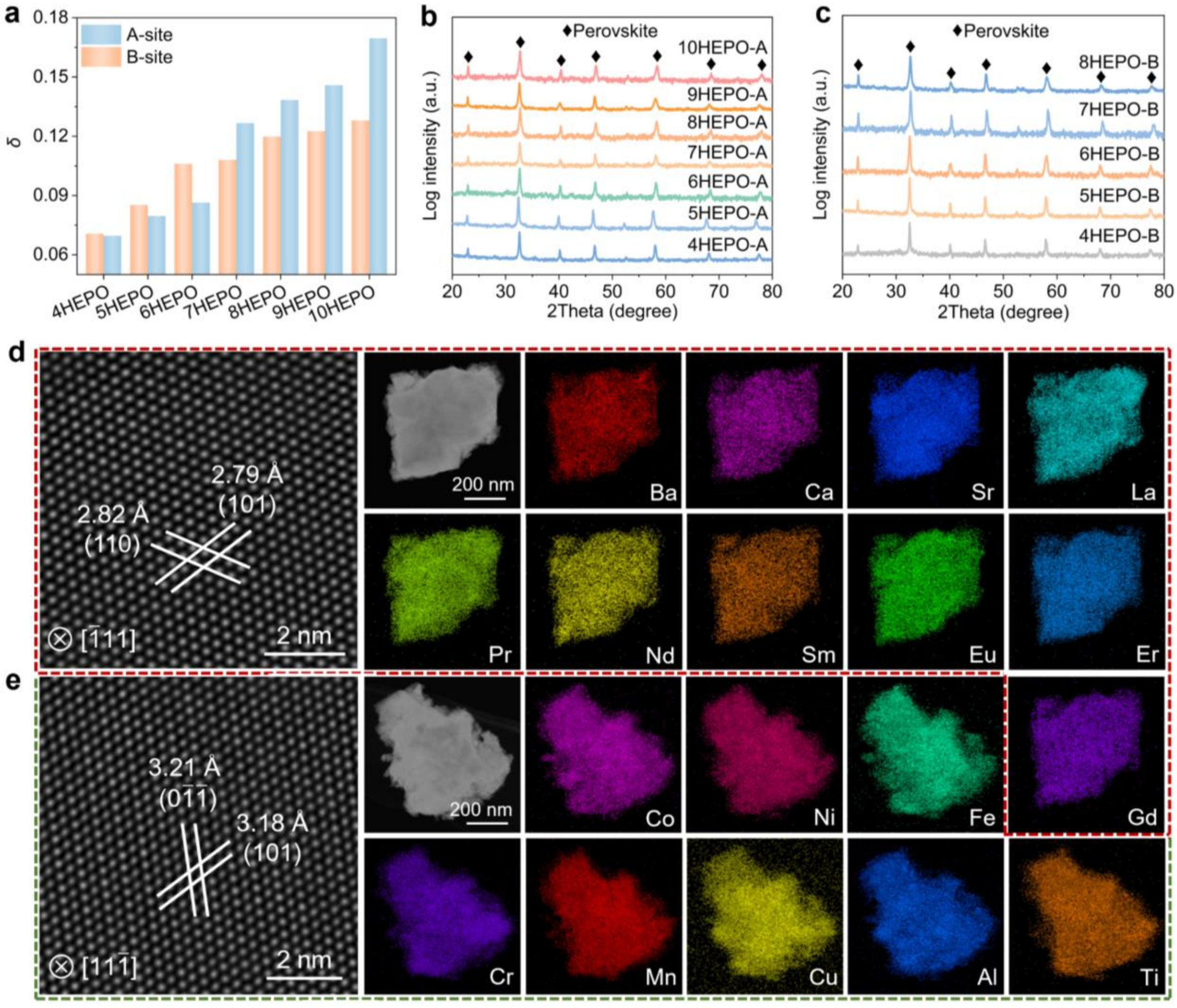

**Fig. 1 Theoretical prediction and experimental synthesis of HEPO-A and HEPO-B samples.** (**a**) $\delta$ of a series of HEPOs. XRD patterns of (**b**) 4–10HEPO-A and (**c**) 4–8HEPO-B samples. Atomic-resolution HAADF-STEM images and STEM-EDS maps of representative (**d**) 10HEPO-A and (**e**) 8HEPO-B samples.

To further validate the structure of the as-synthesized HEPO samples, high-angle annular dark-field (HAADF) imaging was performed via scanning transmission electron microscopy (STEM). Using 10HEPO-A samples as an example, the HAADF-STEM image displays the atomic columns of the B-site cations with adjacent column spacings of 2.82 Å and 2.79 Å, corresponding well to the (110) and (101) planes of the cubic perovskite structure, respectively, along the $[\bar{1}11]$ crystallographic direction (Fig. 1d). Energy-dispersive spectroscopy (EDS) mapping confirms the uniform distribution of all A-site elements, including Ba, Ca, Sr, La, Pr, Nd, Sm, Eu, Er, and Gd, throughout the particles. For the B-site series, the HAADF-STEM image of representative 8HEPO-B samples exhibits clear lattice spacings of 3.21 Å and 3.18 Å, aligning with the $(0\bar{1}\bar{1})$ and (101) planes, respectively, along the $[11\bar{1}]$ zone axis. EDS maps in Fig. 1e also verify the uniform dispersion of the constituent B-site elements (Co, Ni, Fe, Cr, Mn, Cu, Al, and Ti). Collectively, these structural and compositional characterizations provide robust evidence for the successful synthesis of a full series of both A-site and B-site HEPO samples.

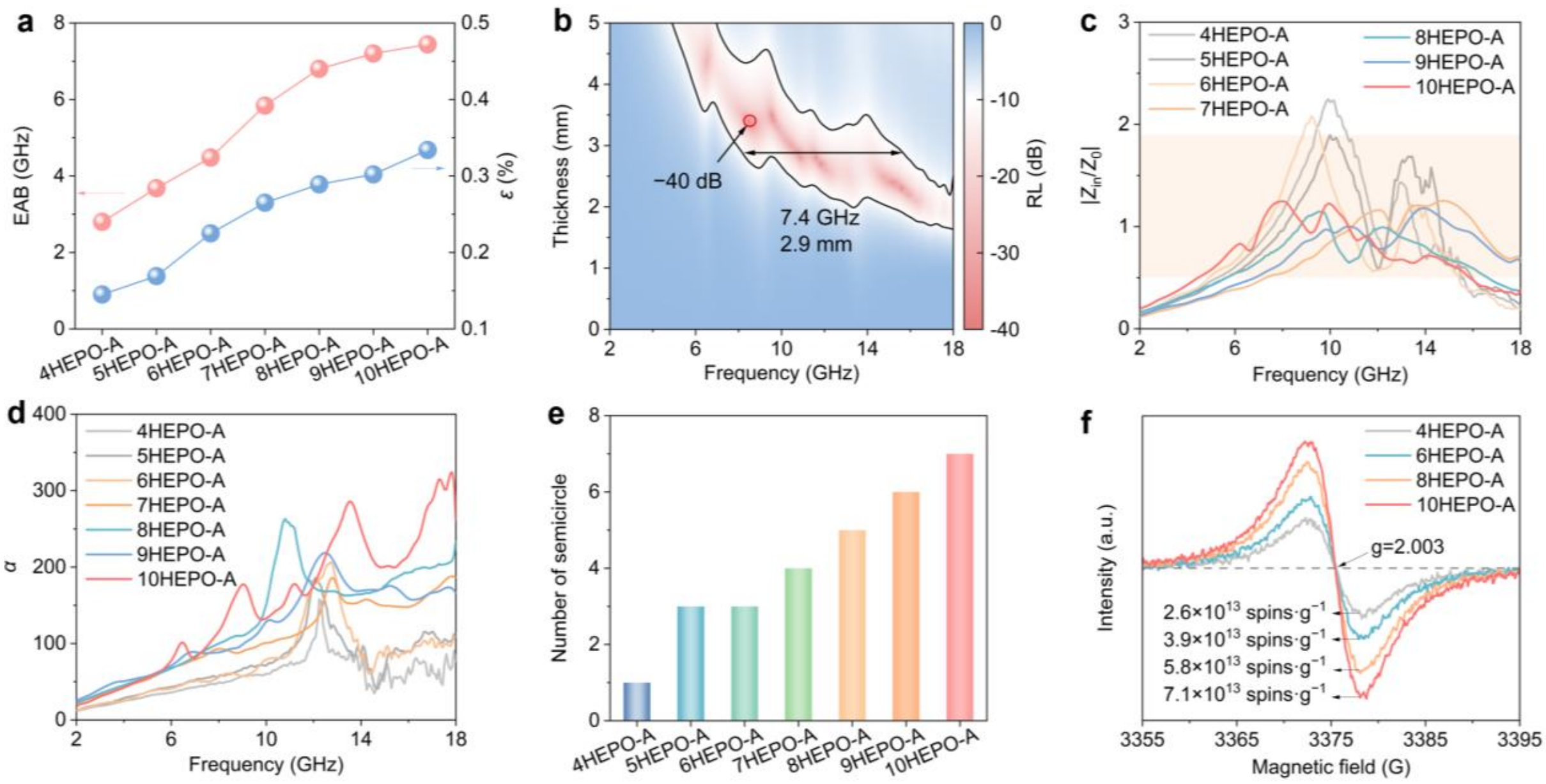

**Fig. 2 EMW absorption performance and mechanisms of 4–10HEPO-A samples**. (**a**) EAB and *ε*. (**b**) RL diagram of 10HEPO-A samples. (**c**) $|Z_{in}/Z_0|$. (**d**) *α*. (**e**) Number of Cole–Cole semicircles. (**f**) EPR quantification of oxygen-vacancy concentrations.

To pinpoint the distinct role of A-site high-entropy configuration, we subsequently examined the EMW absorption properties of the HEPO-A series. As exhibited in Fig. 2a and Fig. S3, the EAB monotonically increases with A-site complexity, peaking at 7.4 GHz with a minimum reflection loss ($RL_{min}$) of −40 dB for 10HEPO-A (Fig. 2b). Also, the lattice strain (*ε*) quantified by the full width at half maximum (FWHM, Fig. S4)[31] displays an upward trend from 0.14 to 0.28 across the 4- to 10-element series (Fig. 2a). These consistent trends imply a strong correlation between EMW absorption and lattice distortion in HEPO-A. To reveal the underlying mechanism responsible for the broadening of EAB, we first analyzed impedance matching, a prerequisite for EMWs to enter absorbers rather than being reflected. In this case, impedance matching is considered satisfactory when $|Z_{in}/Z_0|$ falls within the range of 0.5–1.9 (corresponding to 90% EMW energy absorption), where $Z_{in}$ and $Z_0$ denote the normalized input impedance and the free-space impedance, respectively. As shown in Fig. 2c and Fig. S5, the $|Z_{in}/Z_0|$ values of 4–10HEPO-A samples fluctuate between 56.2% and 71.1%, showing no direct correlation with EAB. This decoupling indicates that impedance matching is not the key factor governing the EMW absorption performance of the HEPO-A series.

Beyond impedance matching, EMW dissipation is equally fundamental. The attenuation constant (*α*), an indicator of dissipation capability, is found to increase monotonically with increasing number of constituent elements (Fig. 2d), demonstrating an intrinsic correlation between EMW dissipation capability and lattice distortion. To further identify the dominating factors for the dissipation enhancement in the HEPO-A series, we next investigated the EM parameters involving complex permittivity ($\varepsilon_r=\varepsilon'-j\varepsilon''$) and permeability ($\mu_r=\mu'-j\mu''$). As shown in Fig. S6 and S7, both $\varepsilon'$ and $\varepsilon''$ rise markedly upon incorporating more cation species, underscoring lattice distortion—induced by the size mismatch among constituent cation atoms—serves as a key contributor to the dielectric loss. To further probe the origin of the enhanced dielectric loss, we analyzed Debye relaxation processes via Cole–Cole plots (Fig. 2e). Among the samples, 10HEPO-A exhibits the most semicircles, indicating the most significant polarization relaxation (Fig. S8). Additionally, the low-frequency tail observed in Fig. S8g points to charge migration or hopping in the 10HEPO-A lattice, revealing that conduction loss adds a non-negligible dissipation contribution. This extra conduction loss is primarily attributed to the increased $\sigma$ (Fig. S9). Electron paramagnetic resonance (EPR) spectra in Fig. 2f corroborate that this $\sigma$ increase arises from enhanced oxygen-vacancy concentrations (g=2.003[32,33]), which rise substantially from 2.6 $\times 10^{13}$ spins$\cdot$g$^{-1}$ to 7.1$\times 10^{13}$ spins$\cdot$g$^{-1}$ as the number of elements increases from 4 to 10 in the HEPO-A series. Besides the dielectric parameters, the magnetic parameters $\mu'$ and $\mu''$ both oscillate around 1 and 0 for 4–10HEPO-A samples, respectively, confirming marginal magnetic loss (Fig. S6).

Having realized the dominant role of dielectric loss in enhancing the EMW absorption properties of A-site HEPOs, we turned our attention to the B-site high-entropy configuration. In keeping with the A-site trend, the EAB of the HEPO-B series also broadens progressively as the elemental diversity increases, accompanied by a concurrent increase in $\varepsilon$ (Fig. 3a). Fig. 3b presents the RL diagram of 8HEPO-B with the maximum number of elements, achieving a substantial EAB of 6.5 GHz at a low thickness of 2.3 mm (RL diagrams for the 4–7HESO-B samples are shown in Fig. S10). $|Z_{in}/Z_0|$ patterns exhibit that impedance matching progressively improves from 4- to 7HEPO-B samples, but worsens slightly at 8HEPO-B (Fig. S11). Such a non-monotonic variation trend implies that impedance matching alone cannot account for the overall EAB enhancement or, at minimum, is not the primary driver. For the EM parameters, as shown in Fig. 3c (derived from Fig. S12), the monotonic increase in both $\varepsilon'$ and $\varepsilon''$ aligns with the EAB broadening, corroborating the critical role of compositional complexity-induced lattice distortion in governing the dielectric loss of the HEPO-B series. In contrast to A-site engineering, the B-site substitution involves magnetic transition metals including Fe, Co, and Ni. Nevertheless, $\mu'$ is observed to vary slightly within the range of 0.96–1.1, while $\mu''$ remains essentially constant at approximately 0, indicating that magnetic loss contributes minimally to the EMW absorption properties of HEPO-B. To substantiate this limited magnetic influence, we performed density of states (DOS) calculations.[34] The spin-polarization in the DOS diagnoses the HEPO-B as magnetic (Fig. 3d). However, the inclusion of more magnetic metallic elements leads to a decrease in the magnetic moment, from 0.81 μB (4HEPO-B) to 0.67 μB (8HEPO-B). This trend is further experimentally verified by their magnetic hysteresis loops: the low-component 4HEPO-B samples display distinct S-shaped signature of ferromagnetism, which gradually weakens with the increasing number of components (Fig. 3e). Since this trend strictly opposes the EAB variation, magnetic loss is confirmed to contribute minimally to EMW attenuation; instead, dielectric loss is the predominant contributor.

Regarding the dielectric loss mechanisms, Cole–Cole plots (Fig. S13) exhibit that the number of semicircles increases with compositional complexity, consistent with the evolutions of $\alpha$ and $\varepsilon$ (Fig. S14). Notably, the pronounced low-frequency tails in 6-, 7-, and 8HEPO-B samples indicate that conduction loss is activated earlier in the B-site HEPOs compared to their A-site counterparts (Fig. S8), potentially underpinning the better EMW absorption performance of the B-site series over the A-site ones at a given number of components. To elucidate the reason for the activation of conduction loss, we comparatively studied the $\sigma$ of 4–8HEPO-A and 4–8HEPO-B samples. Intriguingly, despite the higher oxygen-vacancy concentration that generally facilitates carrier migration in HEPO-A (Fig. 3f), the $\sigma$ of the HEPO-A series is observed to be consistently lower than that of HEPO-B (Fig. 3g). This anomaly indicates that $\sigma$ is governed not solely by the oxygen-vacancy concentration, but by electron transfer through the B-site metal–oxygen network. Based on the crystal field theory,[35,36] $Fe^{2+}$–O–$Fe^{3+}$ channels are critical to facilitating carrier hopping. As revealed by X-ray photoelectron spectroscopy (XPS) analysis (Fig. 3h), lattice distortion due to elemental incorporation in the B-site HEPOs promotes oxygen vacancies. Notably, because of no alternative charge compensation pathway available, the electrons released upon oxygen escape forcibly reduce $Fe^{3+}$ to $Fe^{2+}$.[37] In contrast, although the A-site HEPOs exhibit a higher oxygen-vacancy concentration, the substitution of A-site $La^{3+}$ by lower-valence alkaline-earth metals (e.g., $Sr^{2+}$, $Ca^{2+}$, and $Ba^{2+}$) leads the released electrons to be preferentially consumed for A-site charge compensation, thereby hindering the reduction of $Fe^{3+}$ to $Fe^{2+}$.[38] Consequently, the HEPO-B series exhibits enhanced $\sigma$ even at a relatively low oxygen-vacancy concentration.

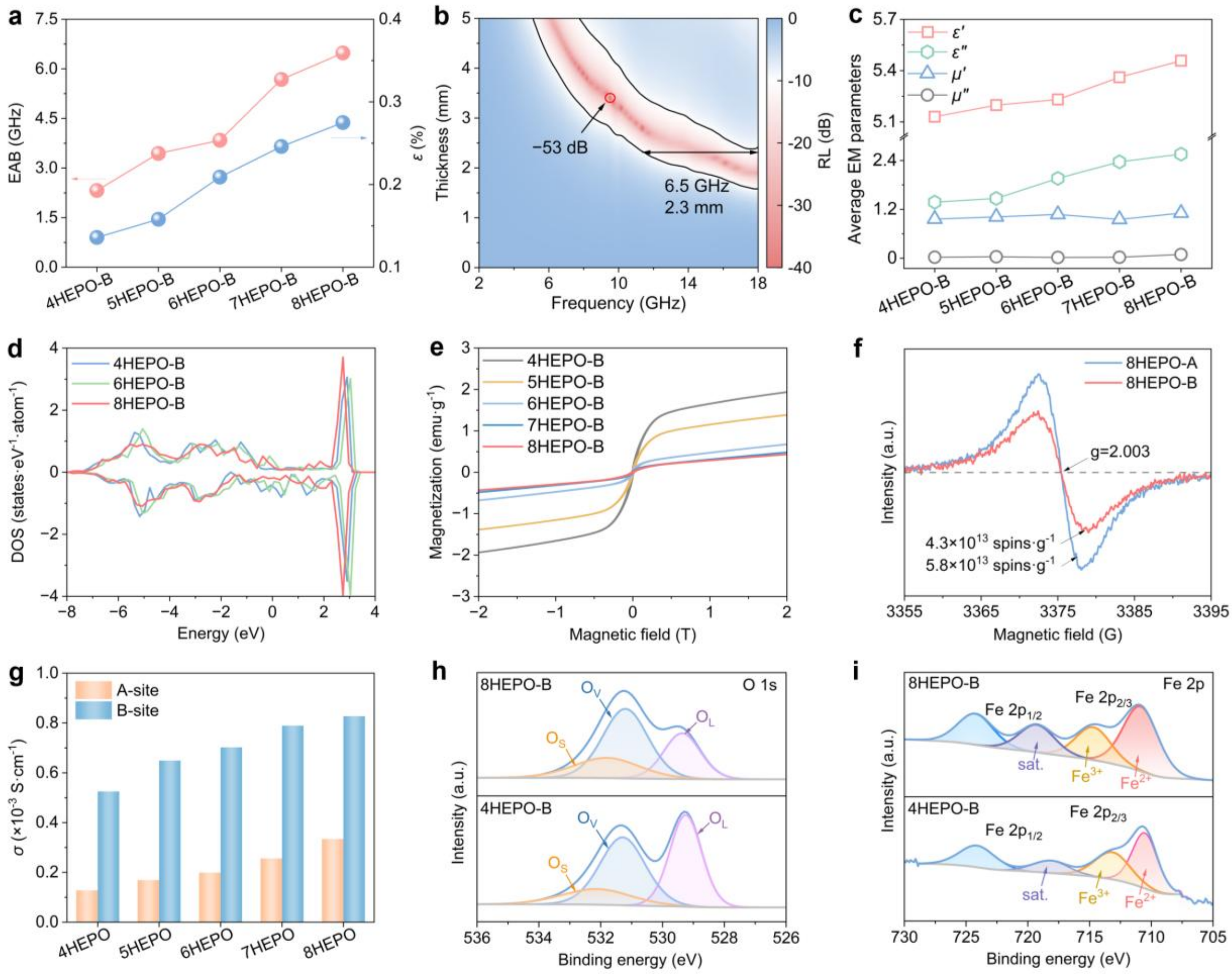

**Fig. 3 EMW absorption performance and mechanisms of 4–8HEPO-B samples.** (**a**) EAB and $\varepsilon$. (**b**) RL diagram of 8HEPO-B samples. (**c**) average EM parameters. (**d**) DOS. (**e**) Magnetic hysteresis loops. (**f**) EPR quantification of oxygen-vacancy concentrations in 8HEPO-A and 8HEPO-B samples. (**g**) $\sigma$. XPS spectra of (**h**) O and (**i**) Fe ions for 4- and 8HEPO-B.

Guided by the key finding that lattice distortion, which can be amplified by elemental incorporation, is paramount to EMW absorption enhancement, we extended this principle by implementing extreme mixing across both A and B sublattices. Using the laser-driven synthesis technique, the designed 9–13HEPO-AB series (detailed compositions are shown in Table S3) with markedly enhanced $\delta$ (Fig. 4a) was synthesized, as confirmed by XRD (Fig. 4b and Fig. S15). Subsequent $\varepsilon$ quantification reveals substantially intensified lattice distortion (Fig. 4c, derived from Fig. S16), corroborating that extreme mixing has been achieved in HEPO-AB. Under such severe lattice distortion, the EAB of the HEPO-AB series is observed to ultimately broaden to 9.2 GHz in 13HEPO-AB, with a low thickness and $RL_{min}$ of 2.7 mm and −50 dB, respectively (Fig. 4d and Fig. S17, derived from Fig. S18). Such outstanding EMW absorption is also largely attributed to the well-optimized impedance matching (Fig. S19). Typically, the Cole–Cole diagram of 13HEPO-AB exhibits multiple distinct Debye-like relaxation semicircles, confirming significant polarization loss.

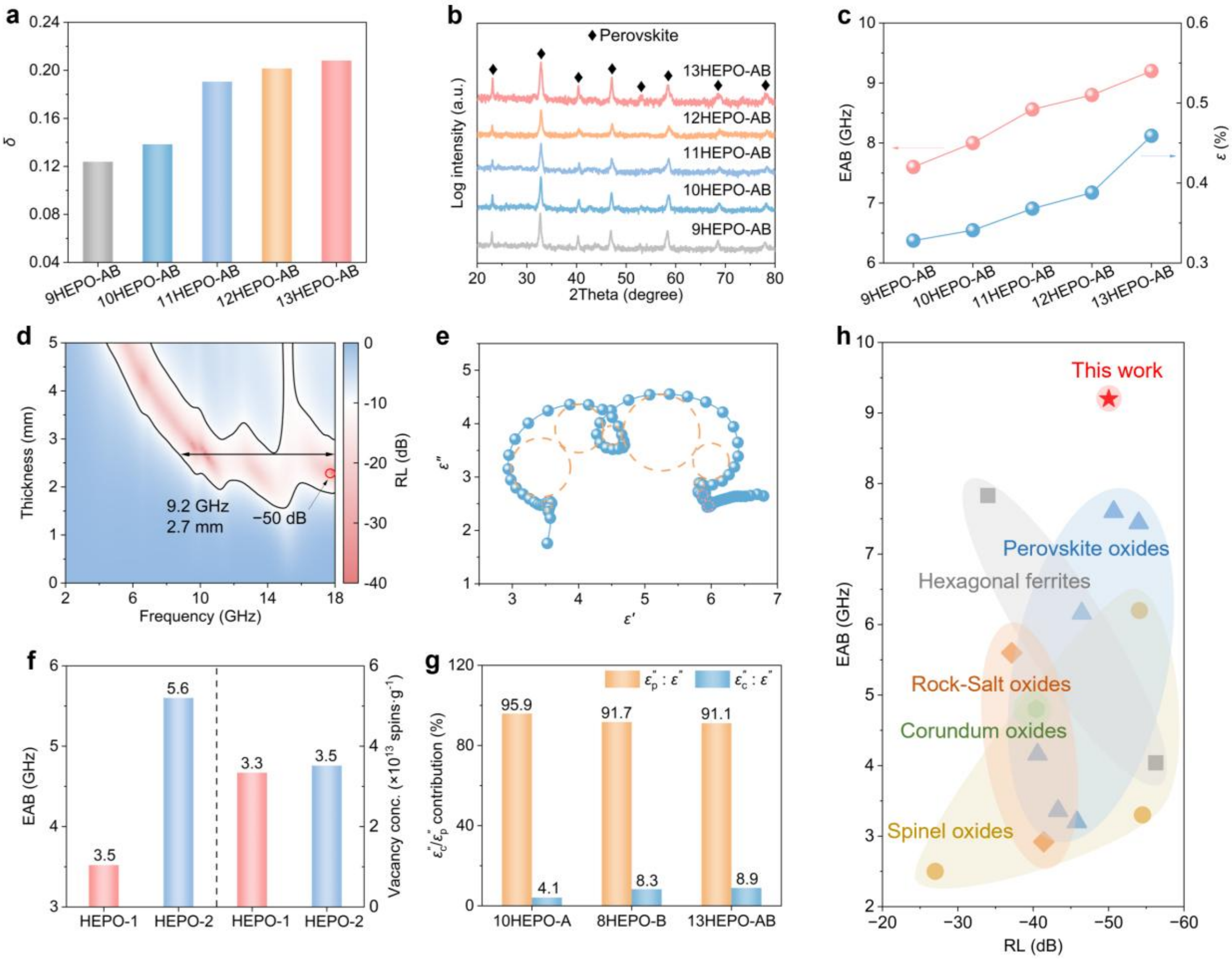

**Fig. 4 EMW absorption performance and mechanisms of 9–13HEPO-AB samples.** (**a**) $\delta$. (**b**) XRD patterns. (**c**) EAB and $\varepsilon$. (**d**) RL diagram and (**e**) Cole–Cole plot of 13HEPO-AB samples. (**f**) EAB comparison between two HEPOs with similar vacancy concentrations. (**g**) Relative contributions of $\varepsilon_p''$ and $\varepsilon_c''$ to $\varepsilon''$. (**h**) Benchmarking the EAB and $RL_{min}$ of 13HEPO-AB against other oxide absorbers.[39-52]

To isolate the polarization loss that arises from oxygen-vacancy-induced dipoles and from atomic-scale interfaces, we conducted a controlled comparative analysis of two synthesizable

HEPOs (phase structures confirmed by XRD in Fig. S20), i.e., La(Fe, Co, Ni, Cr, Mn, Ti, Ga)$O_3$ (HEPO-1) and La(Fe, Co, Ni, Cr, Mn, Ti, Cu)$O_3$ (HEPO-2), with matched vacancy levels ($3.3\times10^{13}$ spin·g$^{-1}$ and $3.5\times10^{13}$ spin·g$^{-1}$, respectively, Fig. S21). Though the vacancy concentrations are essentially the same, the EAB of HEPO-2 (5.6 GHz) appears to be much higher than that of HEPO-1 (3.5 GHz, Fig. 4f, derived from Fig. S22). In contrast, this EAB difference agrees with the $\varepsilon$ variation (0.16% and 0.25%, respectively, Fig. S23). Consequently, it is reasonable to hypothesize that atomic-scale interfaces, rather than oxygen vacancies, dominates the polarization loss that accounts for the majority of the overall dielectric loss contribution ($\varepsilon''$), as evidenced by the calculated relative contributions of polarization loss ($\varepsilon_p''$) and conduction loss ($\varepsilon_c''$, Fig. 4g). Benefiting from the markedly intensified polarization loss, a strikingly broad EAB and an ultralow $RL_{min}$ are thus yielded, underscoring a notable improvement over previously reported oxide absorbers (Fig. 4h and Table S4).

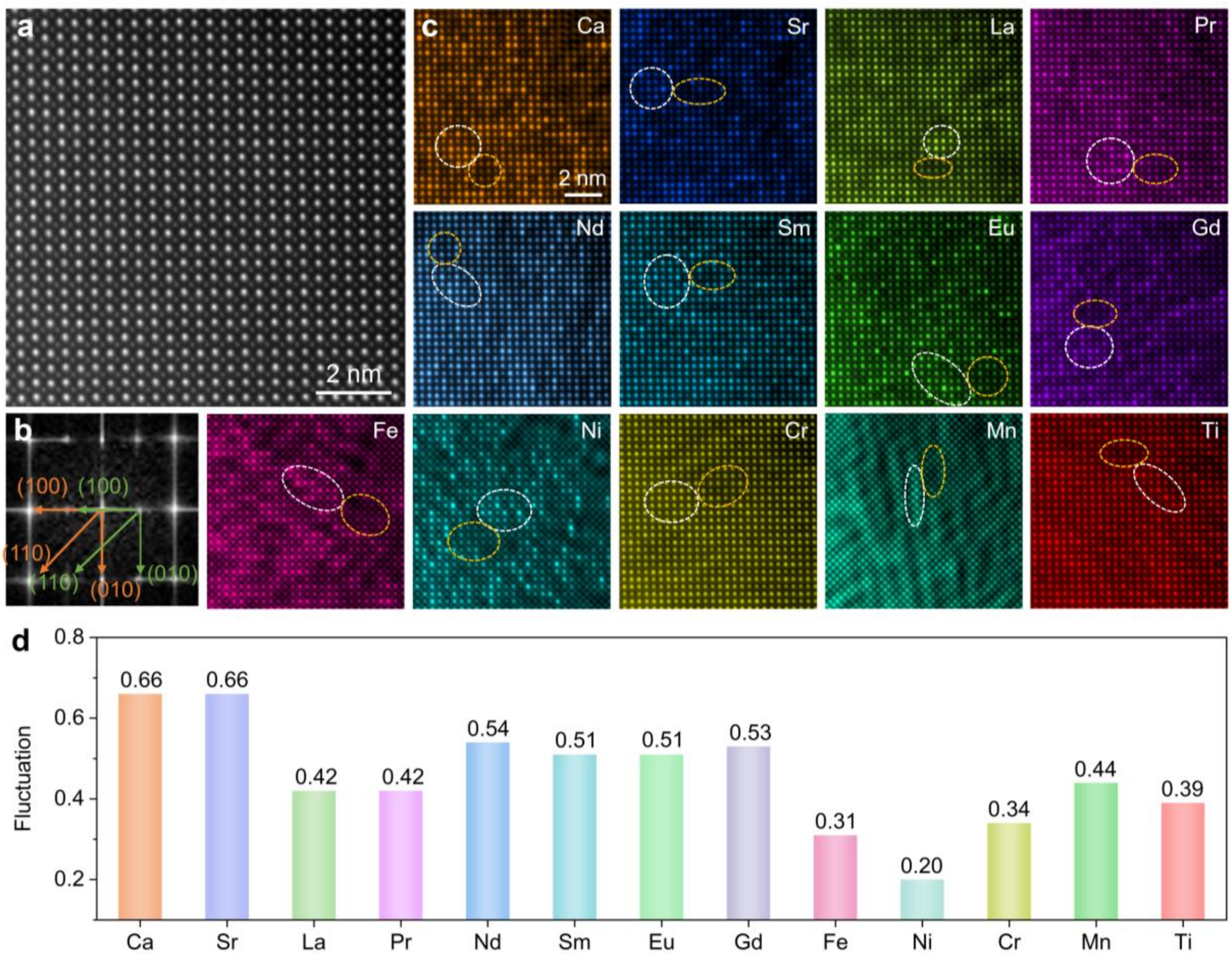

**Fig. 5 Atomic structure of 13HEPO-AB samples.** (**a**) Atomic-resolution HAADF-STEM image. (**b**) FFT pattern of (a). (**c**) Atomic-resolution STEM-EDS maps. (**d**) Quantification of elemental fluctuation for constituent cation elements.

To probe the structural origin of polarization loss at the atomic level, we conducted HAADF-STEM imaging combined with STEM-EDS mapping to directly visualize atomic-scale interfaces. Clearly, two sets of atomic columns with identical *d*-spacings are clearly identified in Fig. 5a, as also confirmed by the corresponding fast Fourier transform (FFT) pattern (Fig. 5b). The periodic enrichment and depletion of different elements within the atomic columns provide compelling evidence of the intensified lattice distortion and the resultant proliferation of atomic-scale interfaces (Fig. 5c), the origin of atomic-scale interfacial polarization loss. To quantify these atomic-scale interfaces, we evaluated the local chemical order (LCO) degree of the constituent elements. As shown in Fig. 5d (derived from Fig. S24), the LCO reaches up to 0.66, exceeding those reported in HEBs (0.34[23]) and HECs (0.54[24]). Therefore, the formation of abundant atomic-scale interfaces, promoted by extreme mixing, is the primary driver of the markedly enhanced EMW absorption performance.

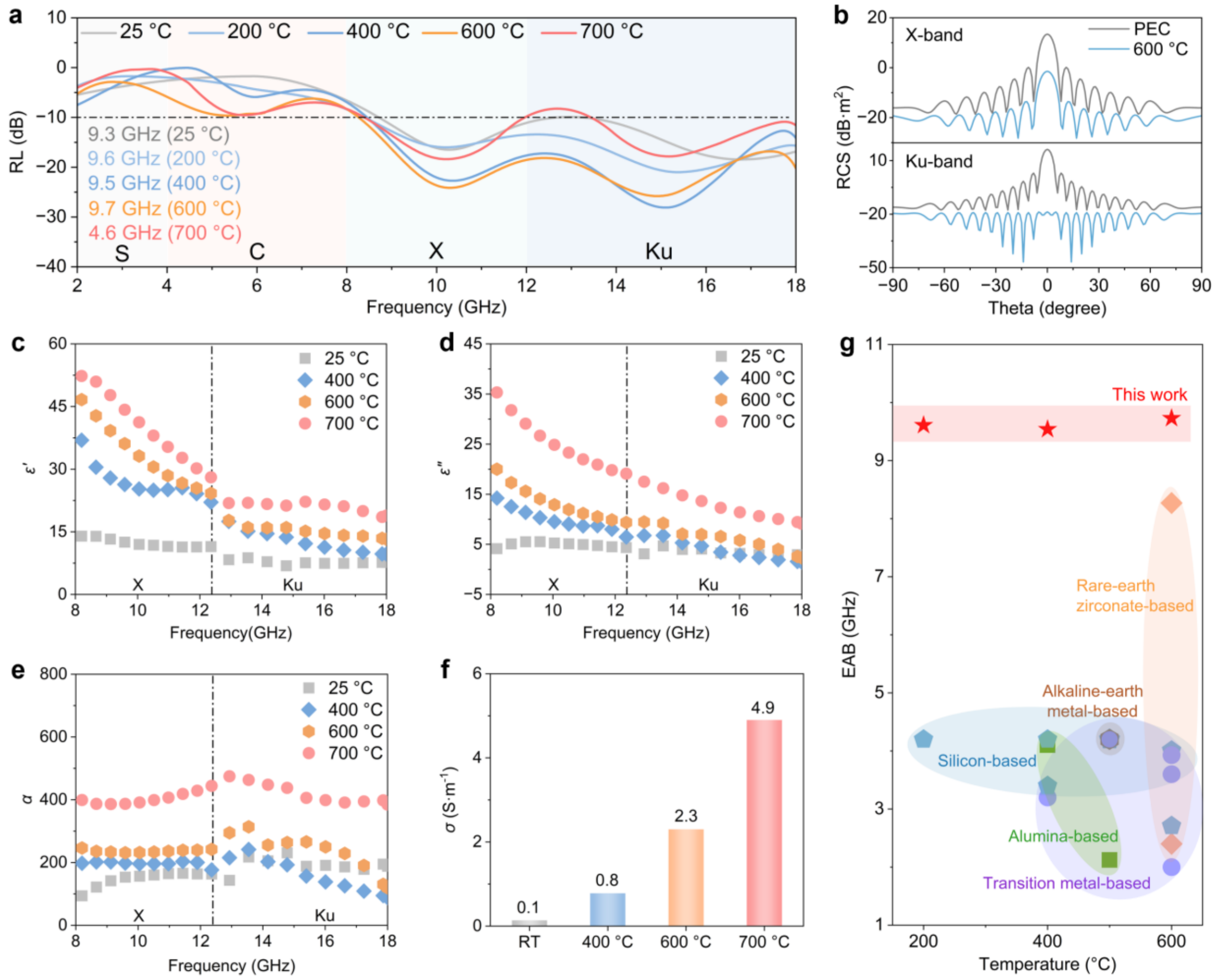

**Fig. 6 High-temperature EMW absorption performance of 13HEPO-AB samples.** (**a**) RL diagram measured from RT to 700 °C by the arch method. (**b**) RCS curves for the PEC and the one coated with 13HEPO-AB at RT and 600 °C in X and Ku B bands. (**c**) $\varepsilon'$, (**d**) $\varepsilon''$, and (**e**) $\alpha$ measured real-time across RT to 700 °C by the waveguide method. (**f**) $\sigma$ across RT to 700 °C. (**g**) Benchmarking 13HEPO-AB against other absorbers in terms of EAB and operating temperature, including silicon-based,[53-58] rare-earth zirconate-based,[18,59] alkaline-earth metal-based,[60] alumina-based,[61,62] and transition metal-based oxides.[43,63-66]

Finally, to thoroughly assess high-temperature applicability, we systematically evaluated the EM absorption performance of 13HEPO-AB at varying temperatures. Real-time arch method measurements show that the as-fabricated 13HEPO-AB plates maintain a broad EAB of 9.3 GHz from RT (25 °C) to 600 °C, with a slight fluctuation of 0.4 GHz (~4% error, Fig. 6a). To further showcase practical potential, we conducted radar cross section (RCS) simulations. Results show that coating a pure perfect electric conductor (PEC) with 13HEPO-AB can markedly suppress radar scattering up to 600 °C, reducing the RCS from 13.4 dB·m$^2$ to −1.5 dB·m$^2$ in the X band and from 16.3 dB·m$^2$ to −18.7 dB·m$^2$ in the Ku band (Fig. 6b). However, as the temperature increases to 700 °C, the EAB drops to 4.6 GHz. To elucidate the mechanism responsible for this EAB drop, we systematically analyzed the dielectric parameters measured via the waveguide method. As shown in Fig. 6c and d, both $\varepsilon'$ and $\varepsilon''$ increase with rising temperature, indicating concurrently enhanced EM wave storage and dissipation capabilities. These enhancements are primarily attributed to the thermally activated hopping conduction, which significantly enhances $\sigma$ (Fig. 6e)—directly elevating $\varepsilon''$ through the dominant conduction-loss contribution and boosting $\varepsilon'$ via hopping-carrier-enhanced polarizations. However, it is noteworthy that the thermally enhanced $\sigma$ (Fig. 6f), arising from increased hopping carrier concentration, results in a severe impedance mismatch, giving rise to the almost total reflection of incident EMWs (Fig. S25–S27). This prevents EM waves from entering the bulk absorber, resulting in the observed decay in the absorption performance at 700 °C. Ultimately, we benchmarked the EAB of as-developed 13HEPO-AB against other oxide absorbers up to 600 °C, including silicon-based,[53-58] rare-earth zirconate-based,[18,59] alkaline-earth metal-based,[60] alumina-based,[61,62] and transition metal-based oxides,[43,63-66] highlighting a significant EM absorption advancement through the extreme mixing strategy (Fig. 6g and Table S5).

**Conclusions**

In summary, through systematic cation modulation of the A and B sublattices, we demonstrate that the EMW absorption performance of HEPOs is governed neither by magnetic loss nor by atomic vacancy-related conduction loss or dipole polarization loss. Instead, it is critically dominated by lattice-distortion-driven atomic-scale interfacial polarization loss—a mechanism that can be maximally amplified through extreme mixing. Guided by this fundamental insight, we further maximize cation incorporation using a laser-driven controllable synthesis technique, successfully developing 13HEPO-AB samples with 8 and 5 cation elements at the A- and B-sites, respectively. Remarkably, the 13HEPO-AB samples exhibit a ultrabroad EAB of 9.3 GHz at RT, which remains 100% fully stable up to 600 °C. At 700 °C, however, the EAB is found to decline to 4.6 GHz. This EAB deterioration is primarily attributed to the thermally increased $\sigma$ due to an enhanced population of hopping carriers, consequently degrading impedance matching that inhibits the effective entry of EMWs. This work develops a broadband high-temperature HEPO absorber with the temperature-stable EAB up to 600 °C by implementing the extreme mixing strategy.

## Methods

### Sample synthesis

The starting materials for the synthesis included commercial rare-earth oxides ($La_2O_3$, $Nd_2O_3$, $Sm_2O_3$, $Pr_6O_{11}$, $Eu_2O_3$, $Er_2O_3$, and $Gd_2O_3$), alkaline-earth metal oxides (BaO, CaO, and SrO), and transition-metal oxides ($Fe_2O_3$, CoO, NiO, $Cr_2O_3$, $MnO_2$, $TiO_2$, and CuO) (purity: 99.9%; particle size: 1–3 μm; Macklin Biochemical Technology Co., Ltd., China). Additionally, $Al_2O_3$ (analytical grade; particle size: >100 mesh; Sinopharm Chemical Reagent Co., Ltd., China) was utilized. Initially, the oxide precursors were accurately weighed according to their designed stoichiometric proportions. The blended powders were then placed inside a jar and subjected to high-energy ball milling for 12 h to achieve uniform mixing. Following this, the mixtures were dried in an oven at 100 °C for 6 h. Once fully dried, the materials were ground into fine precursor powders and tightly packed into a graphite crucible (10 mm in diameter and 3 mm in height). For the rapid synthesis of high-entropy perovskite ceramics, a high-energy laser irradiation system was deployed. Specifically, a fiber continuous 1080 nm-wave laser (RFL-C3000S, Raycus Fiber Laser Technology Co., Ltd., China) served as the primary heating source. The laser beam was adjusted to a diameter of 10 mm via a focusing lens to guarantee an even distribution of energy.[67] During the actual synthesis process, the laser power was maintained at 1000 W with a 100% duty cycle, and the continuous irradiation time was controlled between 3 and 5 s. Ultimately, to obtain uniform particles, the prepared samples were crushed and filtered using a 300-mesh screen.

### Characterization

By quantifying the statistical variation in cation radii, $\delta$ effectively evaluates the degree of lattice distortion and serves as a widely adopted descriptor for structural disorder in HEPOs. Based on the Hume-Rothery solid solution criterion, $\delta$ for A-site cations is defined as follows:

$$\delta_A = \sqrt{\sum_{i=1}^{N} c_i \left(1 - \frac{R_{A_i}}{\left(\sum_{i=1}^{N} c_i R_{A_i}\right)}\right)^2} \tag{1}$$

Where $R_{Ai}$ denotes the ionic radius of the $i$-th cation at the A-site, and $c_i$ represents its molar fraction. The same equation is applied to calculate the $\delta$ value for B-site cations.

The vacancy defects of the prepared HEPO specimens were analyzed via EPR spectroscopy (EMXplus-6/1/P/L, Bruker, USA) across a magnetic-field span from 200 to 6200 G. Additionally, $\sigma$ tests were conducted with a four-point probe resistivity system (280SI, Four Dimensions, USA). Magnetization of the HEPO samples were evaluated utilizing a vibrating sample magnetometer (Dynacool-9, Quantum Design, USA) at an applied magnetic field of 3 T. Phase compositions were determined via X-ray diffraction (X'pert PRO, PANalytical, Netherlands). The collected XRD profiles were evaluated through Rietveld refinement utilizing GSAS-II. The $\varepsilon$ of HEPO powders was computed applying the Williamson-Hall approach based on the subsequent equations:

$$\varepsilon = \frac{D\beta_T \cos\theta - K\lambda}{4D\sin\theta} \tag{2}$$

where $\beta_T$ represents FWHM of the diffraction peak, $\theta$ denotes the Bragg angle, $K$ stands for the shape factor, $\lambda$ indicates the wavelength of the applied X-ray radiation, and $D$ refers to the mean crystallite size.

A Cs-corrected STEM (Spectra 300, Thermo Fisher Scientific, USA) equipped with four Super-X EDS detectors was utilized at 300 kV for atomic-level characterization. Both HAADF imaging (25 mrad convergence angle, 61–200 mrad collection angle) and EDS mapping (25 pm pixel size, 8 μs dwell time, ~30 min total acquisition) were conducted under a constant beam current of 50 pA. Initial EDS spectrum images obtained via Velox software were subjected to non-local principal component analysis for denoising to ensure structural accuracy.[68] Ultimately, atomic site intensities in these refined EDS maps were quantified using CalAtom software, adopting the identical fitting algorithm applied to the HAADF images.[69] EM parameters in the 2–18 GHz range were measured with a vector network analyzer (E5071C, Agilent Technologies, USA). Based on transmission-line theory, the RL and $\alpha$ were calculated using the following equations:[46]

$$RL = 20\log\left|\frac{Z_{\mathrm{in}}-Z_0}{Z_{\mathrm{in}}+Z_0}\right| \tag{3}$$

$$Z_{\mathrm{in}} = Z_0\sqrt{\frac{u_r}{\varepsilon_r}}\tanh\left[j\left(\frac{2\pi f d}{c}\right)\sqrt{u_r\varepsilon_r}\right] \tag{4}$$

$$\alpha = \frac{\sqrt{2}\pi f}{c}\times\sqrt{(\mu''\varepsilon''-\mu'\varepsilon')+\sqrt{(\mu''\varepsilon''-\mu'\varepsilon')^2+(\mu'\varepsilon''+\mu''\varepsilon')^2}} \tag{5}$$

where $f$ represents the EM-wave frequency, $c$ denotes the speed of light, and $d$ indicates the thickness of the testing sample.

Based on the Debye relaxation theory, $\varepsilon'$ can be decomposed into $\varepsilon_c''$ and $\varepsilon_p''$ The contribution of each component was determined by the following formulas:[70]

$$\varepsilon'' = \frac{2\pi f\tau(\varepsilon_s-\varepsilon_\infty)}{1+(2\pi f\tau)^2}+\frac{\sigma}{2\pi f\varepsilon_0} = \varepsilon_p''+\varepsilon_c'' \tag{6}$$

$$\varepsilon_c'' = \frac{\sigma}{2\pi f\varepsilon_0} \tag{7}$$

where $\varepsilon_0$ denotes the vacuum permittivity ($8.85\times10^{-12}$ F·m$^{-1}$).

For high-temperature electromagnetic measurements, the synthesized 13HEPO-AB powders were loaded into graphite dies (60 mm in diameter and 50 mm in height) and consolidated via spark plasma sintering under a biaxial pressure of 35 MPa. The sintering process was conducted at 1500 °C for 15 min in an argon atmosphere, followed by natural cooling to yield the test samples. High-temperature $\varepsilon_r$ was measured via the waveguide method across the X and Ku bands using a vector network analyzer (E5071C, Agilent Technologies, USA) integrated with a heating module. The sintered pellets were machined into specific dimensions: 22.86 mm × 10.16 mm × 2 mm for the X-band and 15.80 mm × 7.90 mm × 2 mm for the Ku-band. During testing, samples were heated at a rate of 10 °C·min$^{-1}$ and maintained for 10 min at each target temperature prior to data collection. Following the GJB 2038A–2011 standard, the RL of plate-shaped 13HEPO-AB samples (180 mm × 180 mm × 4 mm) were assessed between 2 and 18 GHz using the arch method in an anechoic chamber. The sample was positioned at the arc's center and illuminated by a source antenna at a 10° oblique incidence angle, while a second antenna captured the reflected signals. To investigate EM absorption at elevated temperatures (200 °C, 400 °C, 600 °C, and 700 °C), the thermal state was precisely monitored by five thermocouples located at the center and four corners of the plate. At each temperature stage, including RT, the samples were stabilized for 10 min before the RL values were recorded.

CST STUDIO SUITE 2023 software was utilized to simulate the RCS of 13HEPO-AB samples relying on the experimentally acquired permittivity and permeability. The model was constructed as a plate (180 × 180 $mm^2$), consisting of a 1.0 mm-thick PEC substrate and a 1.2 mm-thick absorbing coating layer. The incident angle of the electromagnetic waves varied from −180° to 180°. The RCS values were computed via the subsequent equation:[19]

$$RCS(\mathrm{dB \cdot m^2}) = 10\,log\left(\frac{4\pi S}{\lambda^2}\left|\frac{E_s}{E_i}\right|^2\right) \tag{8}$$

where $S$ represents the area of the conducting plate, $\lambda$ denotes the wavelength of the incident wave, while $E_S$ and $E_i$ stand for the magnitudes of the scattered and incident electric fields, respectively.

**Computation**

Based on the conventional $LaFeO_3$ unit cell, three specific supercell models, 4HEPO-B (1×2×1), 6HEPO-B (1×4×1), and 8HEPO-B (1×6×1), were generated using the special quasi-random structure (SQS) approach,[71] implemented in the Alloy Theoretic Automated Toolkit (ATAT).[72] All density functional theory (DFT) computations were carried out using the Vienna Ab initio Simulation Package (VASP) with spin polarization considered.[73] The interactions between electrons and ions were modeled via projector augmented-wave (PAW) pseudopotentials,[71] while the Perdew-Burke-Ernzerhof functional (PBE) within the generalized gradient approximation (GGA) was adopted to treat the exchange-correlation energy. To properly capture the intricate magnetic ground states inherent in these multi-cation systems, spin polarization was considered in all simulations. A kinetic energy cutoff of 520 eV was chosen for the plane-wave basis. For Brillouin zone integration, a Γ-centered mesh with a k-point spacing of roughly 0.3 $Å^{-1}$ was applied. Geometry relaxations were considered complete when the change in total energy fell below the threshold of $10^{-5}$ eV. Following the structural optimization, DOS was evaluated by employing a more refined k-point grid of 5×3×5.

**Acknowledgments**
Authors acknowledge the financial support from the National Natural Science Foundation of China (Nos. 52572072 and 52472072), Guangzhou Basic and Applied Basic Research Foundation (No. SL2024A04J01220), Guangdong Basic and Applied Basic Research Foundation (No. 2025A1515010644).

**Author contributions**

Y.C designed and supervised the research. Z.F and F.G conducted the experiments. W.W conducted the STEM imaging. Y.L and H.Y performed the theoretical calculations. Y.C, L.Z, Z.F, and W.W analyzed the data and wrote the manuscript. All authors contributed helpful discussion to this work.

**Competing interests**
Authors declare that they have no competing interests.

Supplementary Information

# Extreme mixing enables broadband high-temperature electromagnetic absorption in high-entropy perovskites

Ziyu Feng[1], Fangchao Gu[1], Lei Zhuang[1*], Wu Wang[2*], Hulei Yu[1], Yanhui Chu[1*]

[1]School of Materials Science and Engineering, South China University of Technology, Guangzhou, 510641, China.

[2]College of Integrated Circuits and Optoelectronic Chips, Shenzhen Technology University, Shenzhen, 518118, China.

[*]Corresponding author. lzhuang@scut.edu.cn (L.Z); wangwu@sztu.edu.cn (W.W); chuyh@scut.edu.cn (Y.C).

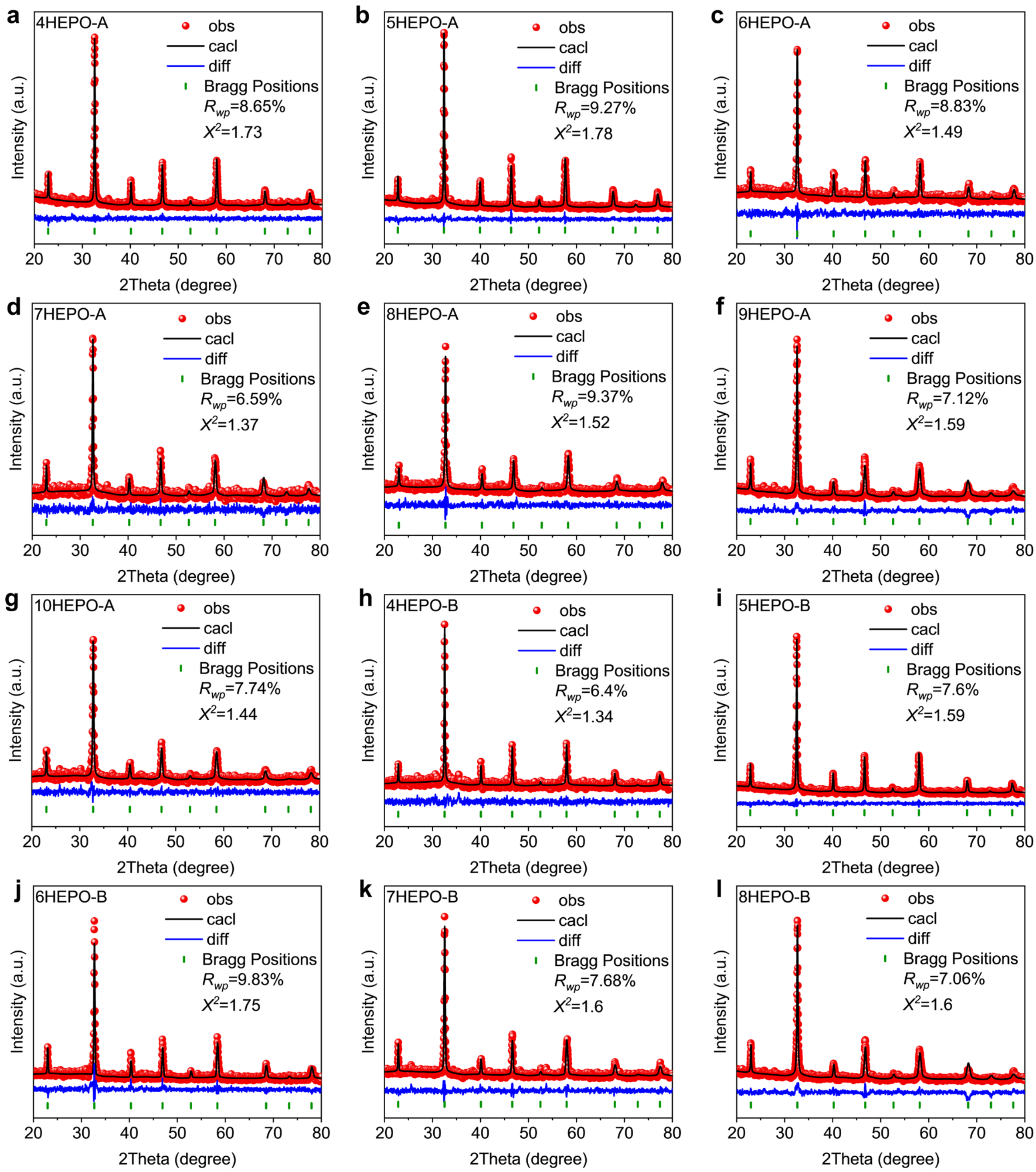

**Figure S1.** XRD Rietveld-refinement profiles of HEPO-A and HEPO-B samples. (a) 4HEPO-A. (b) 5HEPO-A. (c) 6HEPO-A. (d) 7HEPO-A. (e) 8HEPO-A. (f) 9HEPO-A. (g) 10HEPO-A. (h) 4HEPO-B. (i) 5HEPO-B. (j) 6HEPO-B. (k) 7HEPO-B. (l) 8HEPO-B.

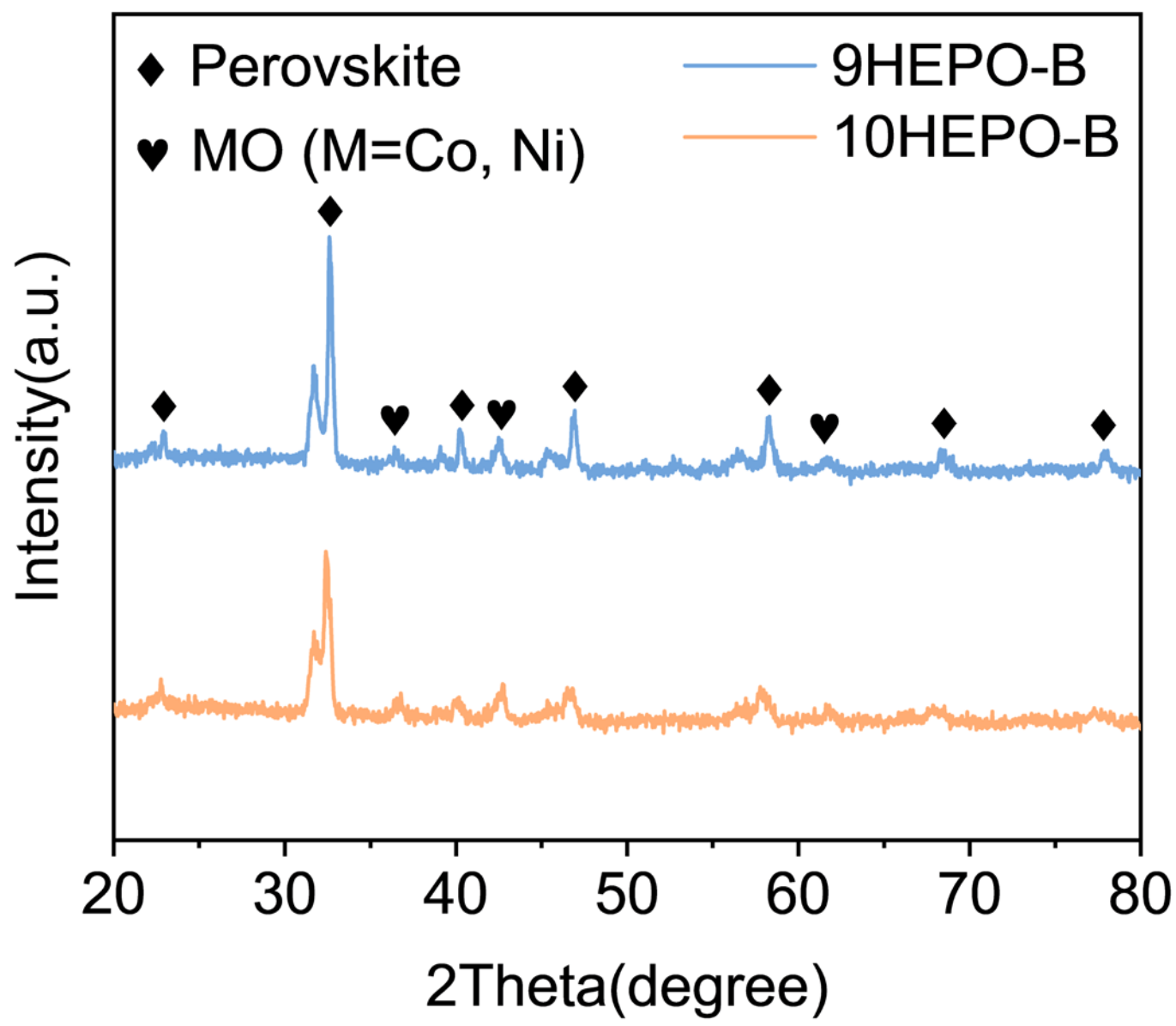

**Figure S2.** XRD patterns of 9HEPO-B and 10HEPO-B samples.

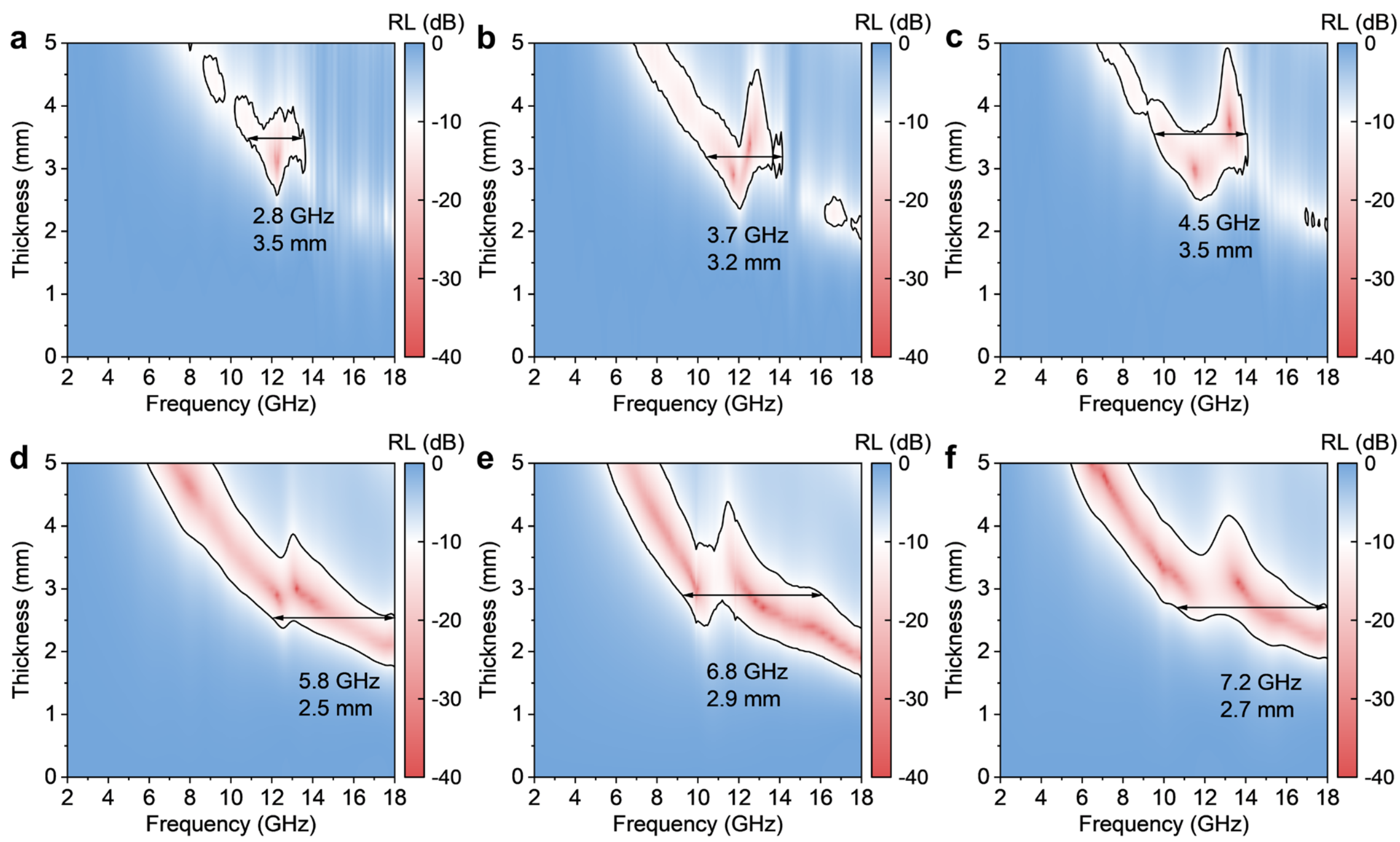

**Figure S3.** EMW absorption performance of HEPO-A samples. (a) 4HEPO-A. (b) 5HEPO-A. (c) 6HEPO-A. (d) 7HEPO-A. (e) 8HEPO-A. (f) 9HEPO-A.

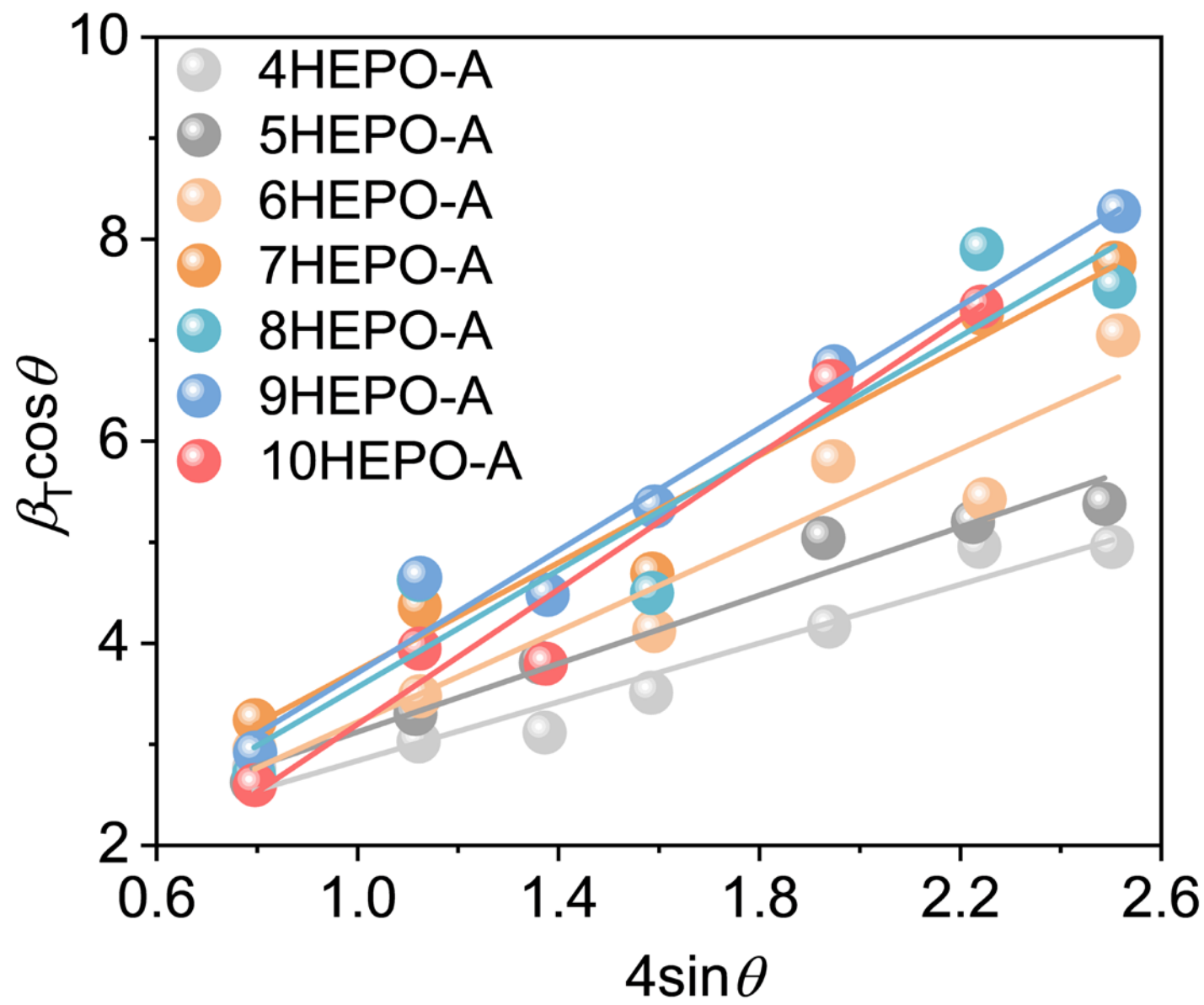

**Figure S4.** $\varepsilon$ of HEPO-A samples.

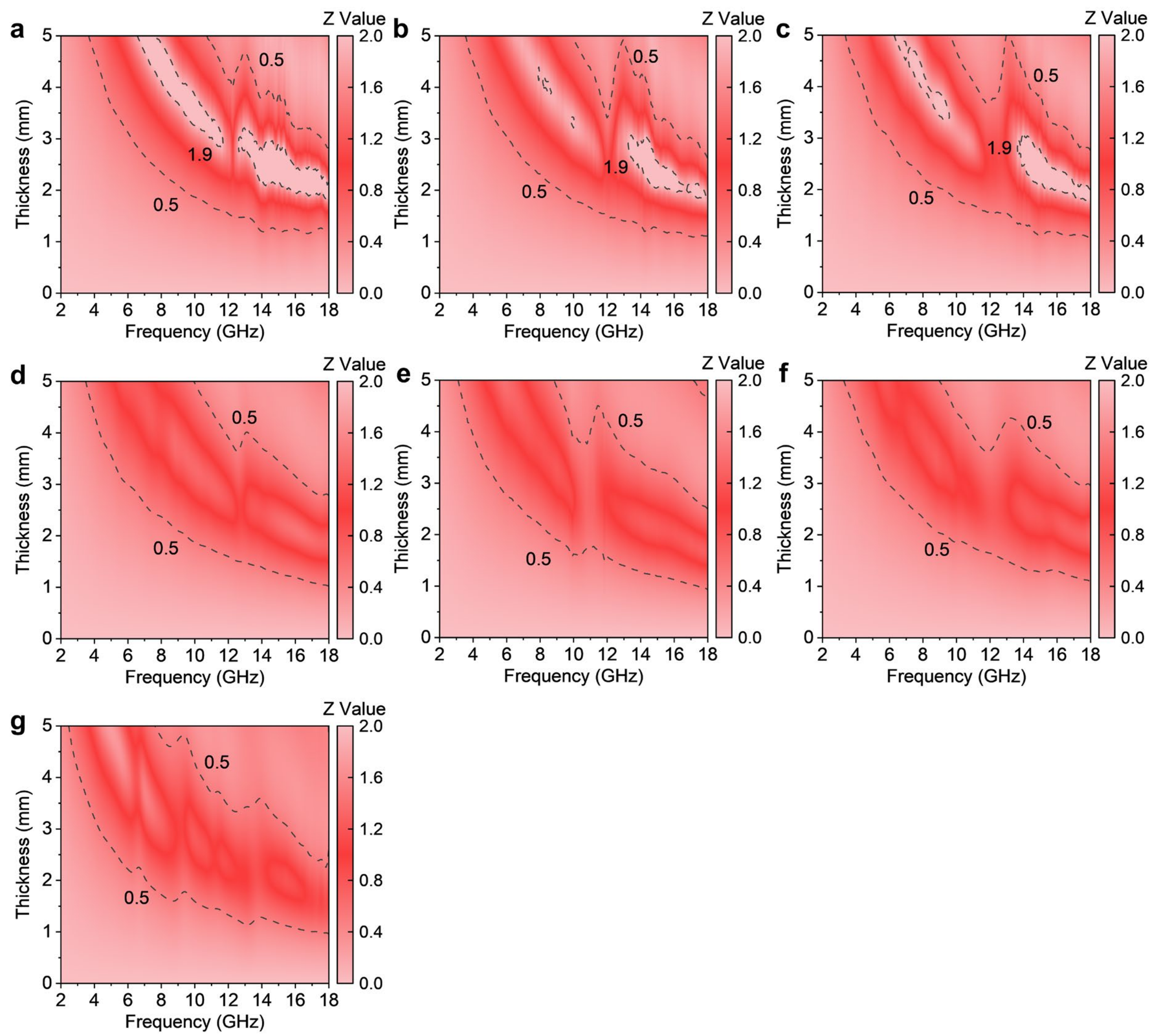

**Figure S5.** $|Z_{in}/Z_0|$ patterns of HEPO-A samples. (a) 4HEPO-A. (b) 5HEPO-A. (c) 6HEPO-A. (d) 7HEPO-A. (e) 8HEPO-A. (f) 9HEPO-A. (g) 10HEPO-A.

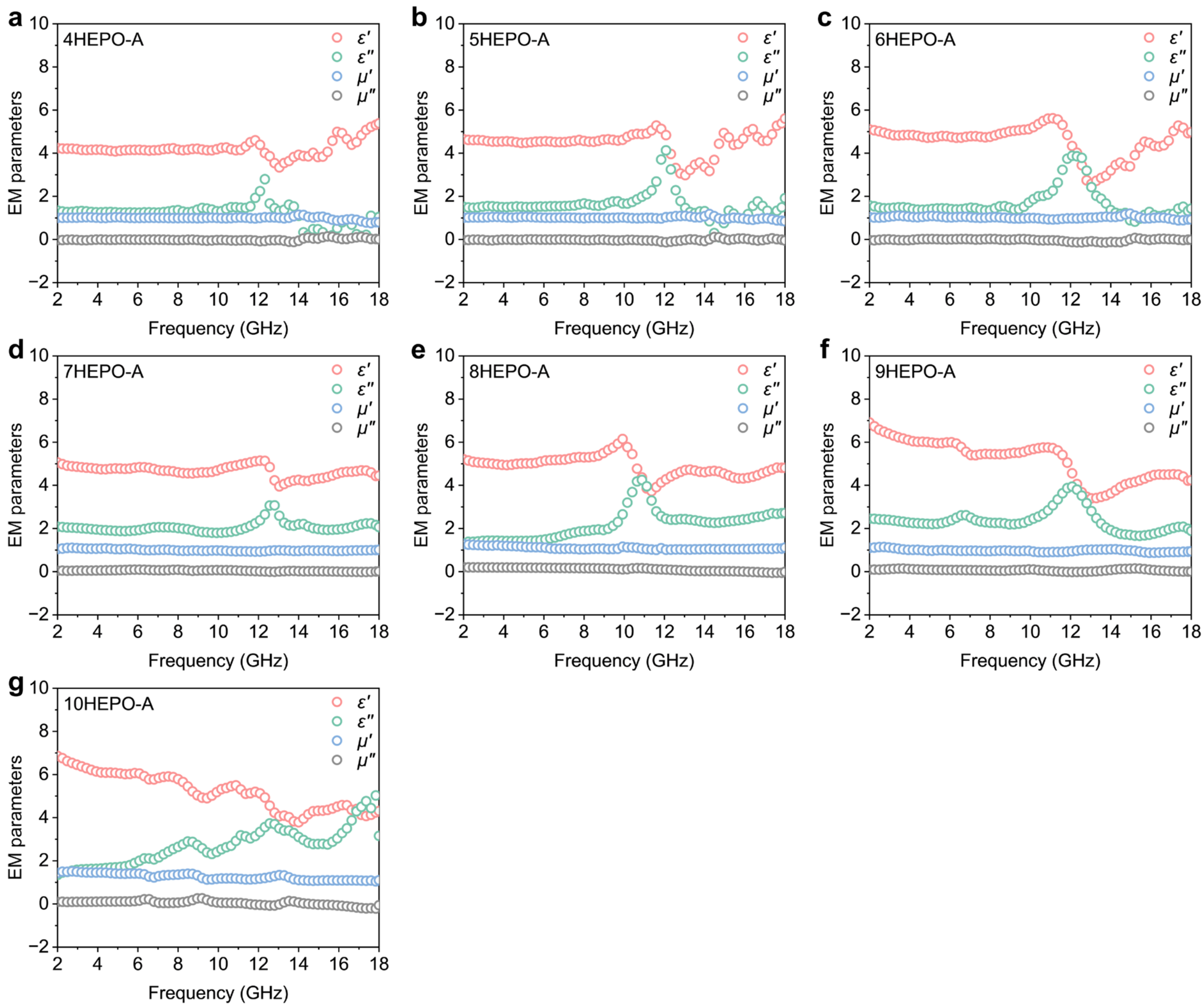

**Figure S6.** EM parameters of HEPO-A samples. (a) 4HEPO-A. (b) 5HEPO-A. (c) 6HEPO-A. (d) 7HEPO-A. (e) 8HEPO-A. (f) 9HEPO-A. (g) 10HEPO-A.

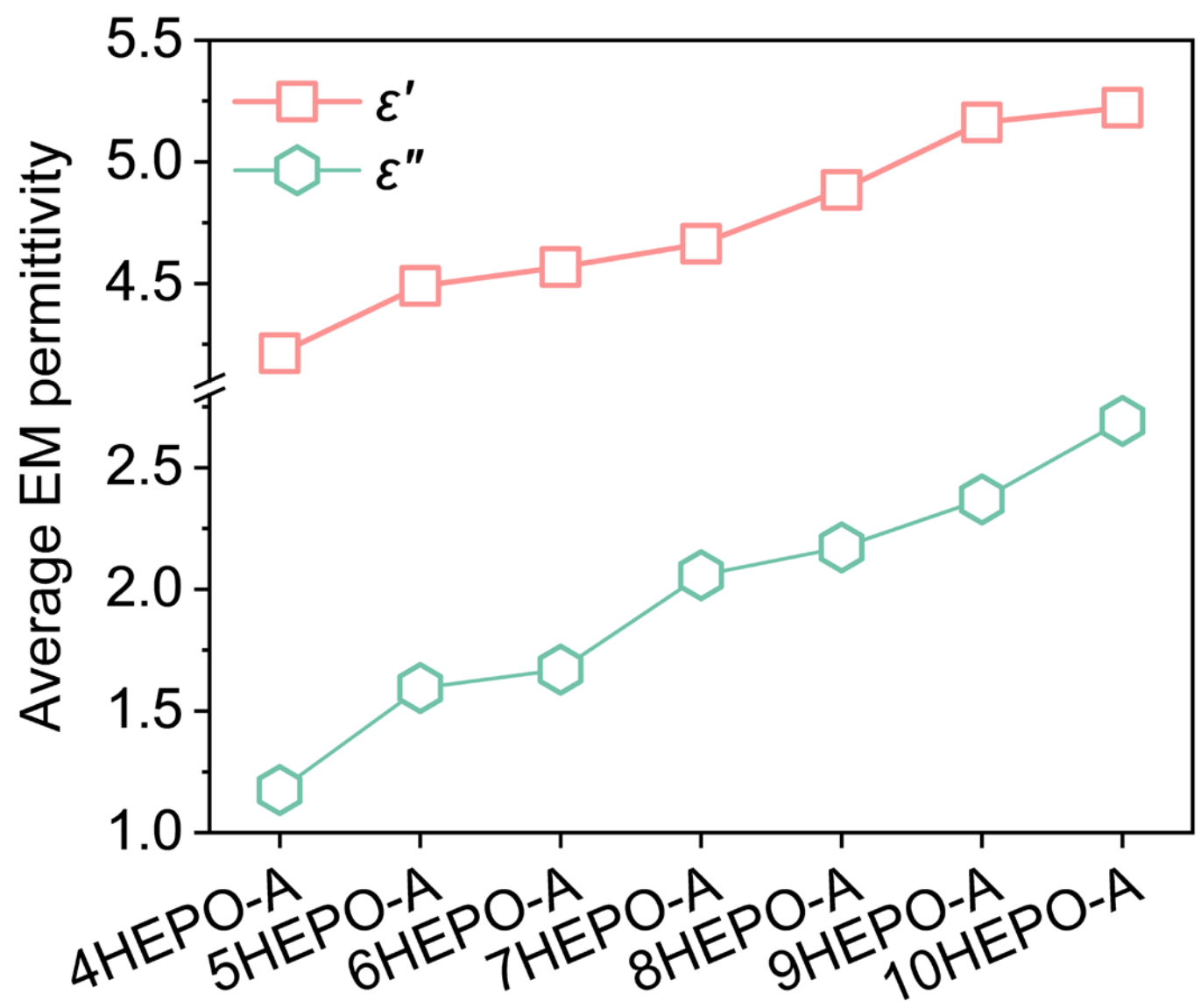

**Figure S7.** Average $\varepsilon'$ and $\varepsilon''$ of HEPO-A samples.

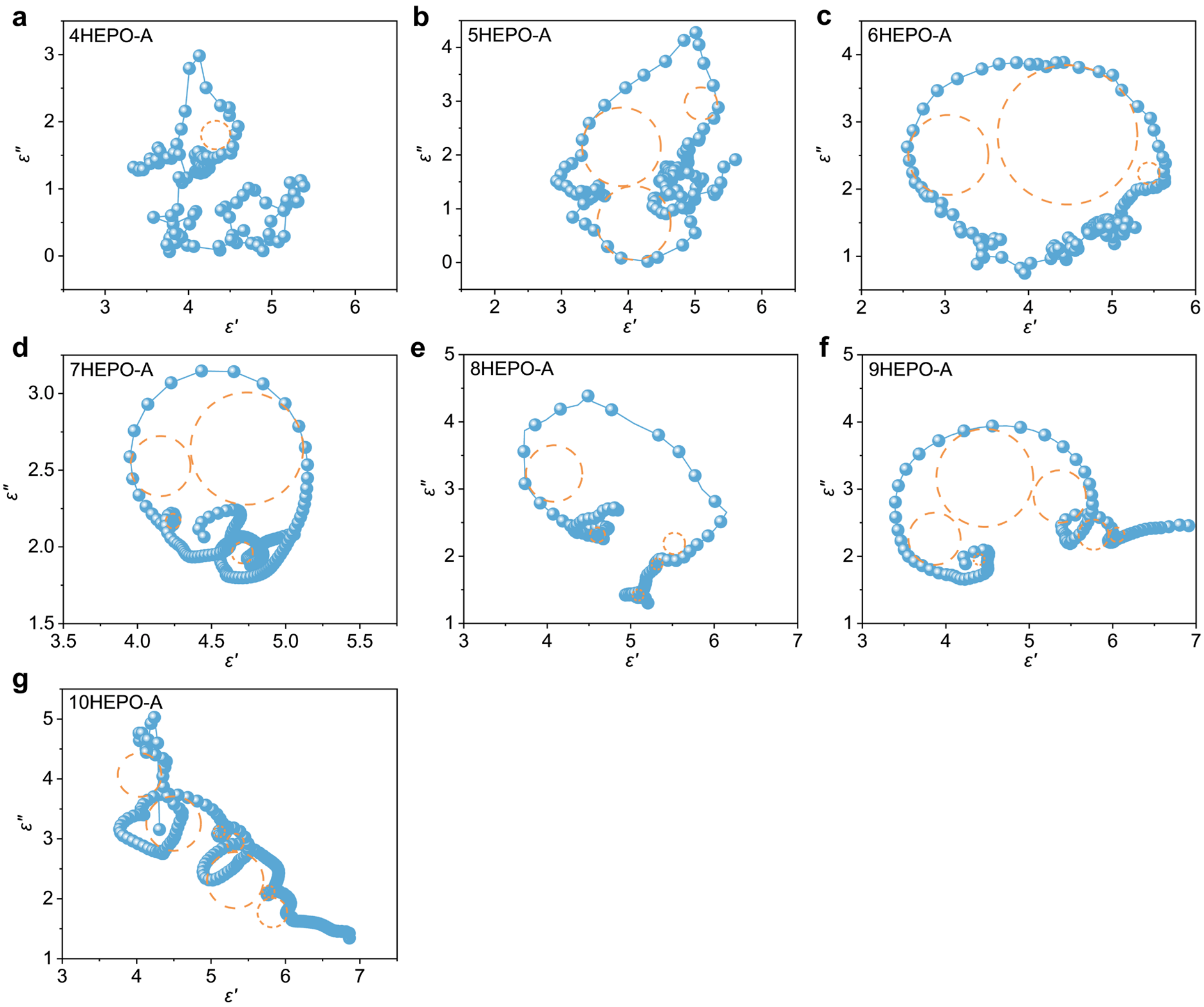

**Figure S8.** Cole–Cole plots of HEPO-A samples. (a) 4HEPO-A. (b) 5HEPO-A. (c) 6HEPO-A. (d) 7HEPO-A. (e) 8HEPO-A. (f) 9HEPO-A. (g) 10HEPO-A.

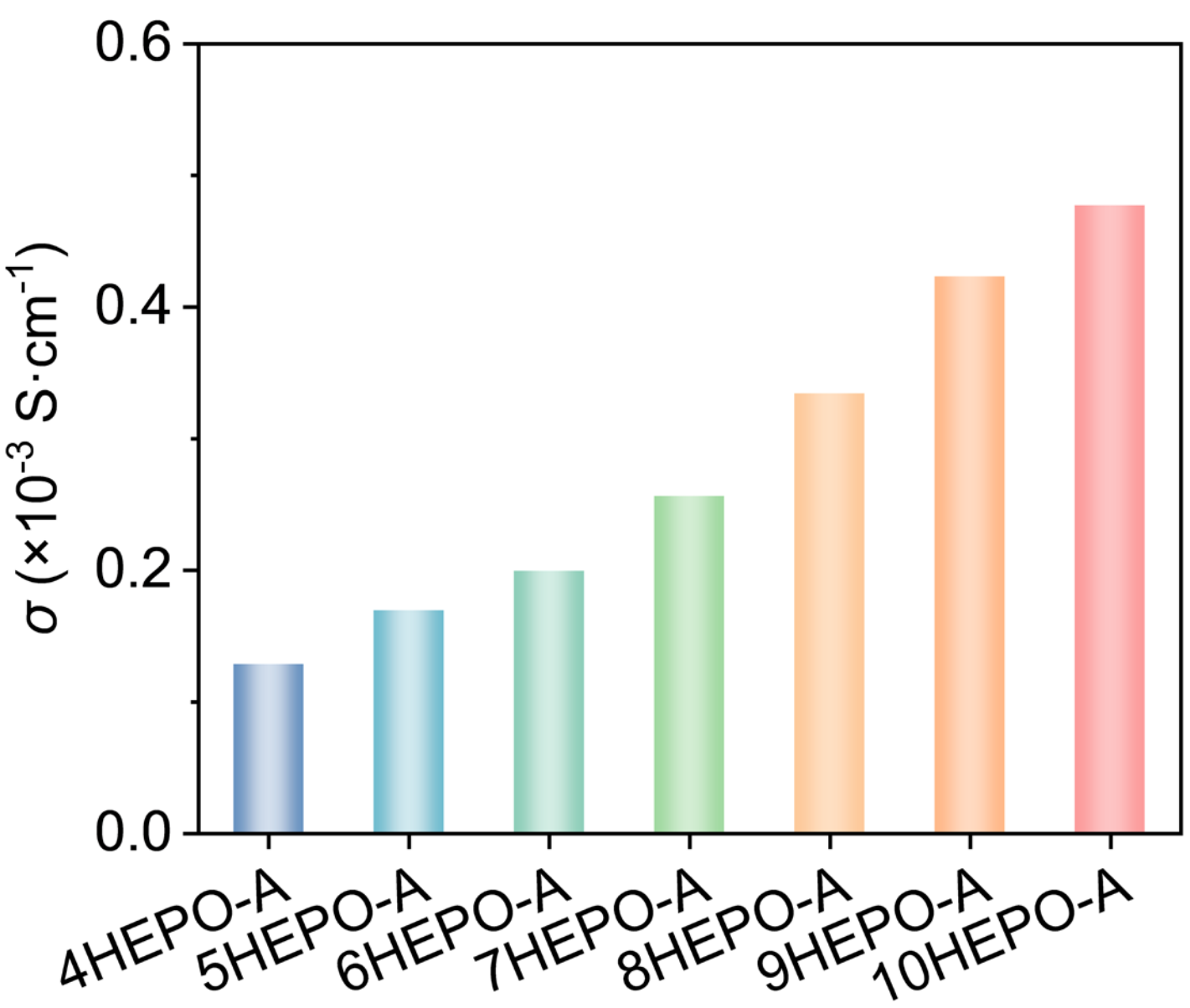

**Figure S9.** Measured $\sigma$ of HEPO-A samples.

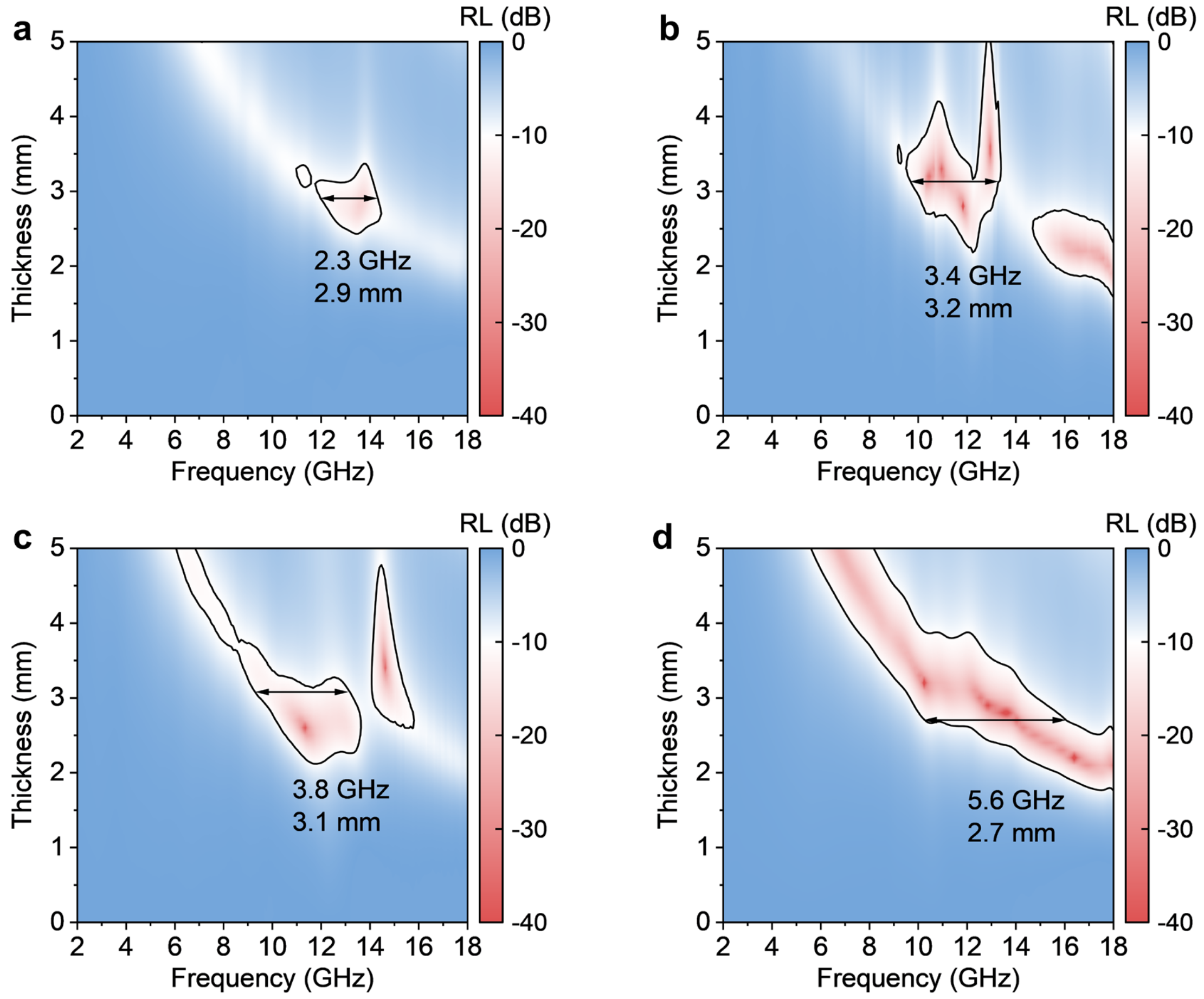

**Figure S10.** EMW absorption performance of HEPO-B samples. (a) 4HEPO-B. (b) 5HEPO-B. (c) 6HEPO-B. (d) 7HEPO-B.

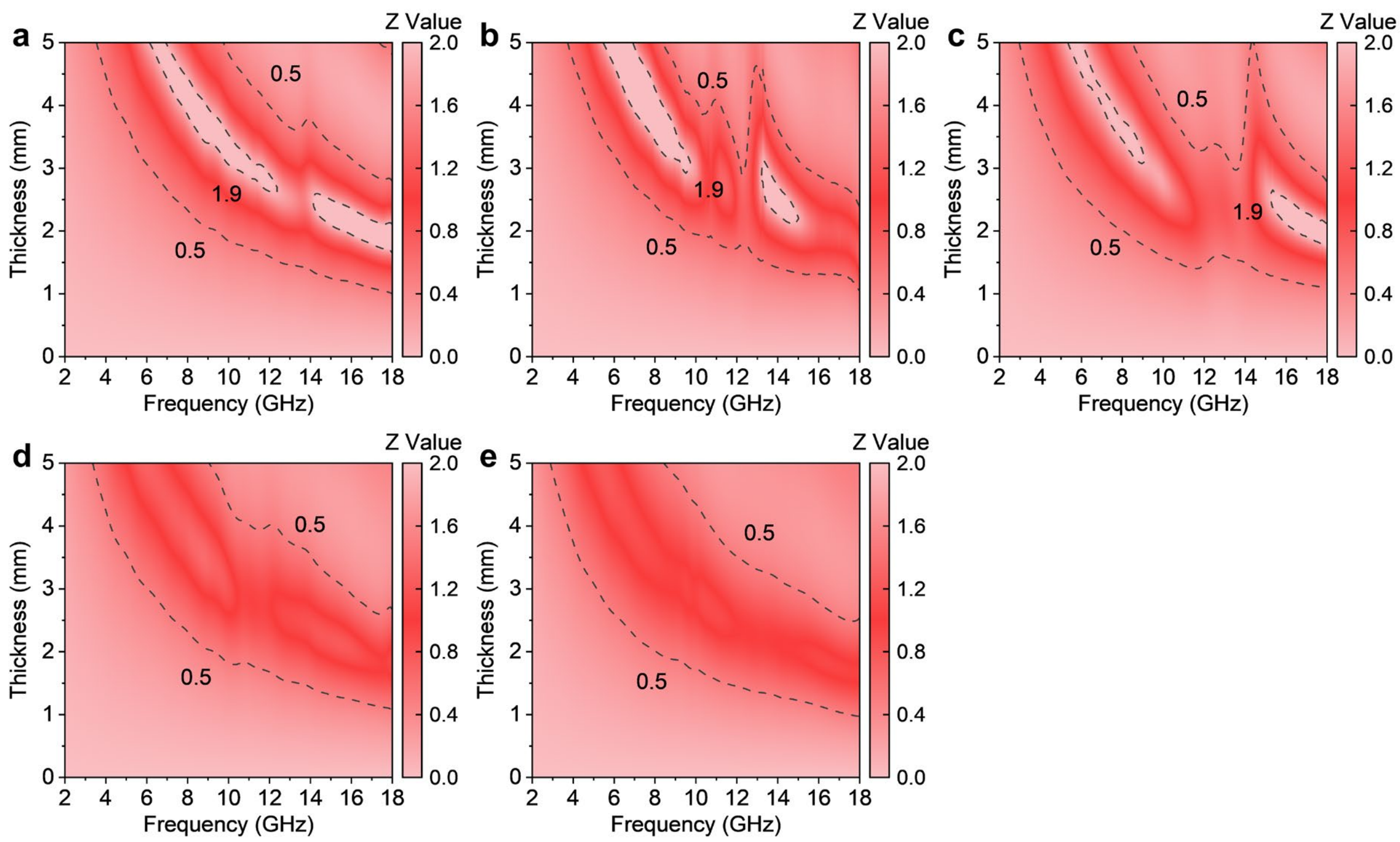

**Figure S11.** $|Z_{in}/Z_0|$ patterns of HEPO-B samples. (a) 4HEPO-B. (b) 5HEPO-B. (c) 6HEPO-B. (d) 7HEPO-B. (e) 8HEPO-B.

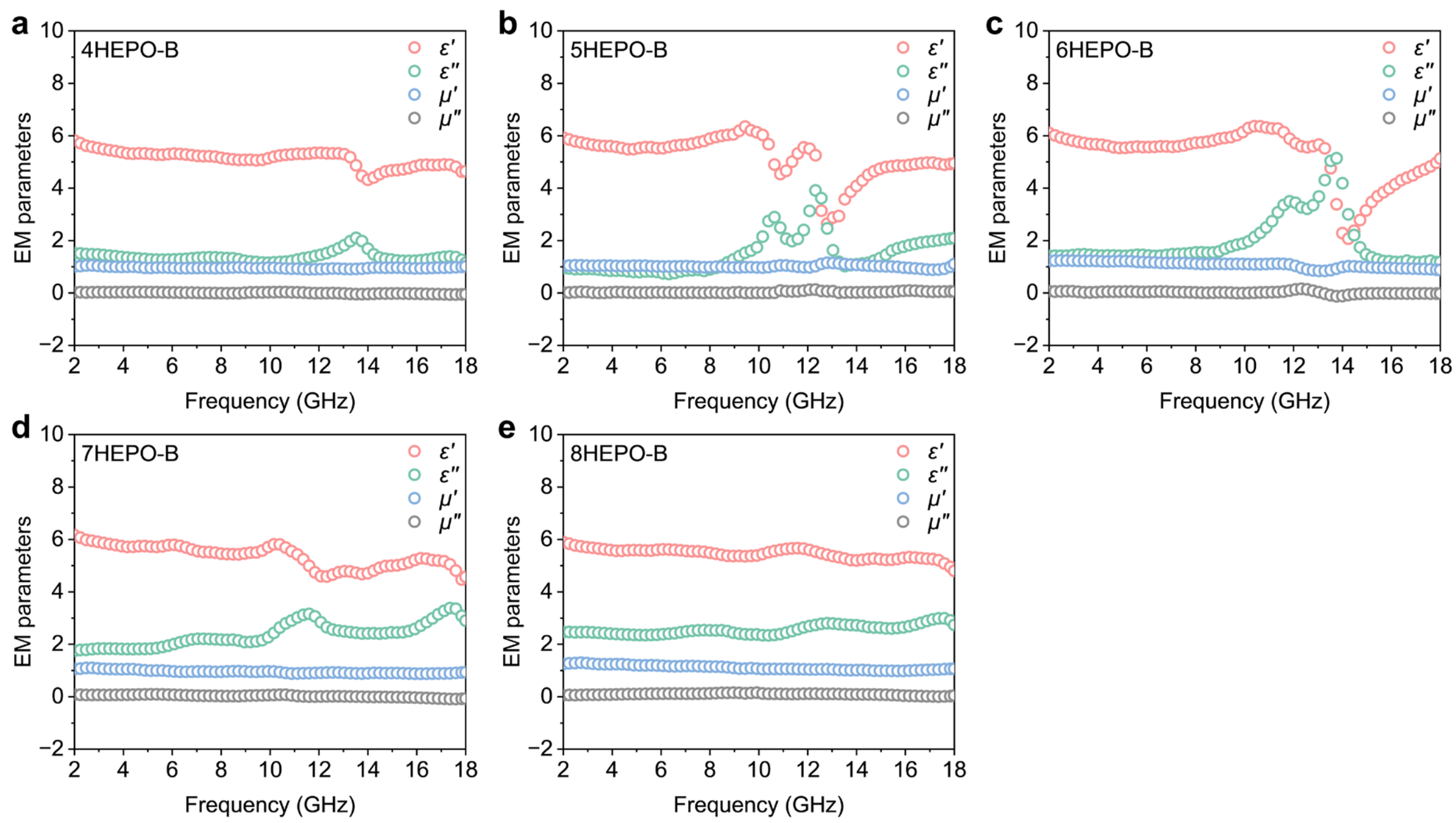

**Figure S12.** EM parameters of HEPO-B samples. (a) 4HEPO-B. (b) 5HEPO-B. (c) 6HEPO-B. (d) 7HEPO-B. (e) 8HEPO-B.

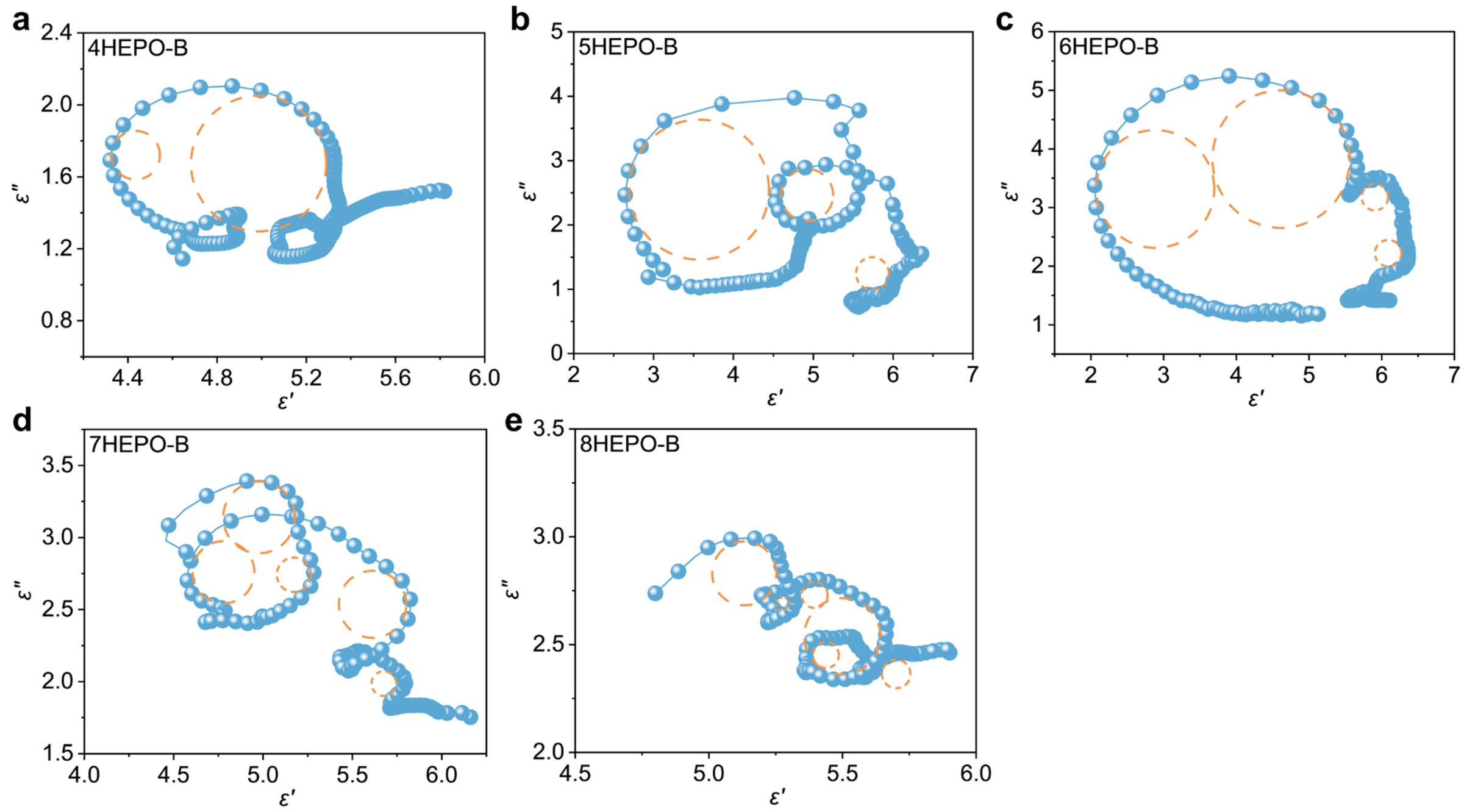

**Figure S13.** Cole–Cole plots of HEPO-B samples. (a) 4HEPO-B. (b) 5HEPO-B. (c) 6HEPO-B. (d) 7HEPO-B. (e) 8HEPO-B.

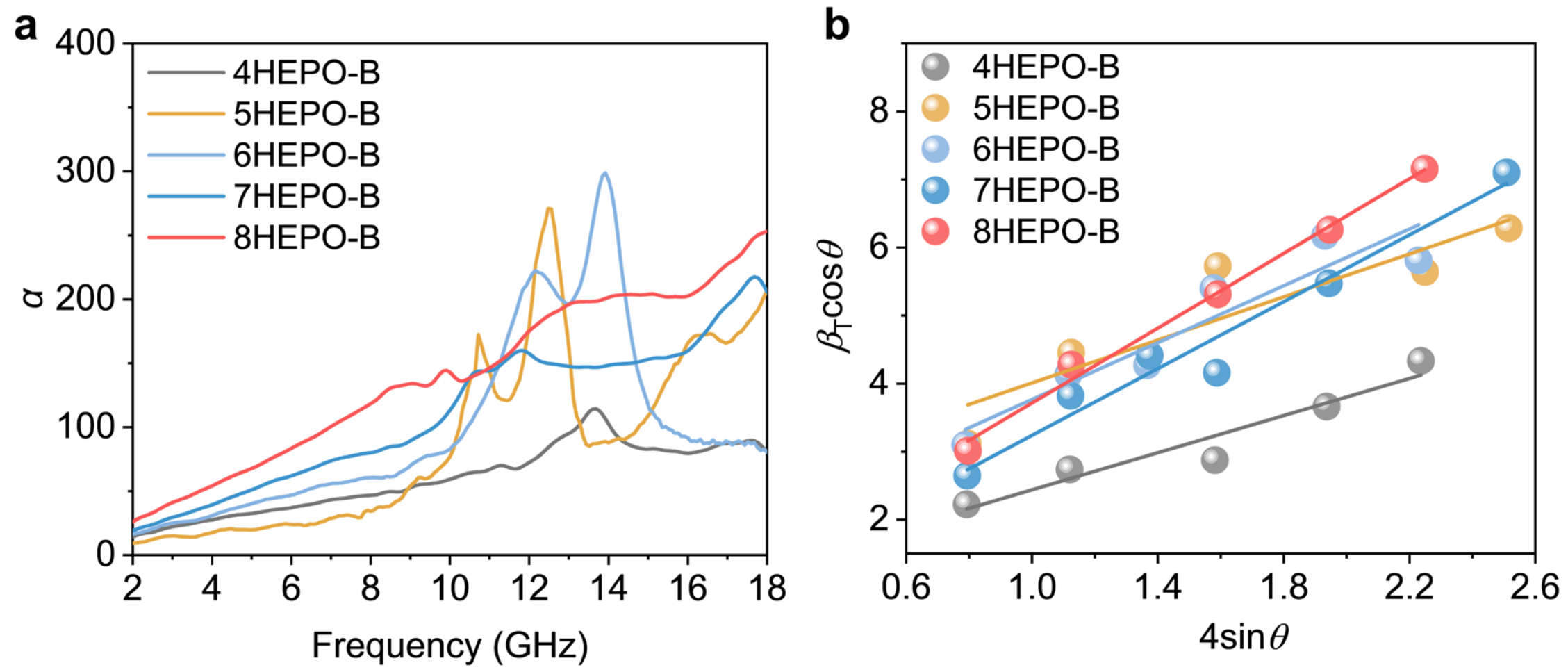

**Figure S14.** (a) $\alpha$ and (b) $\varepsilon$ of HEPO-B samples.

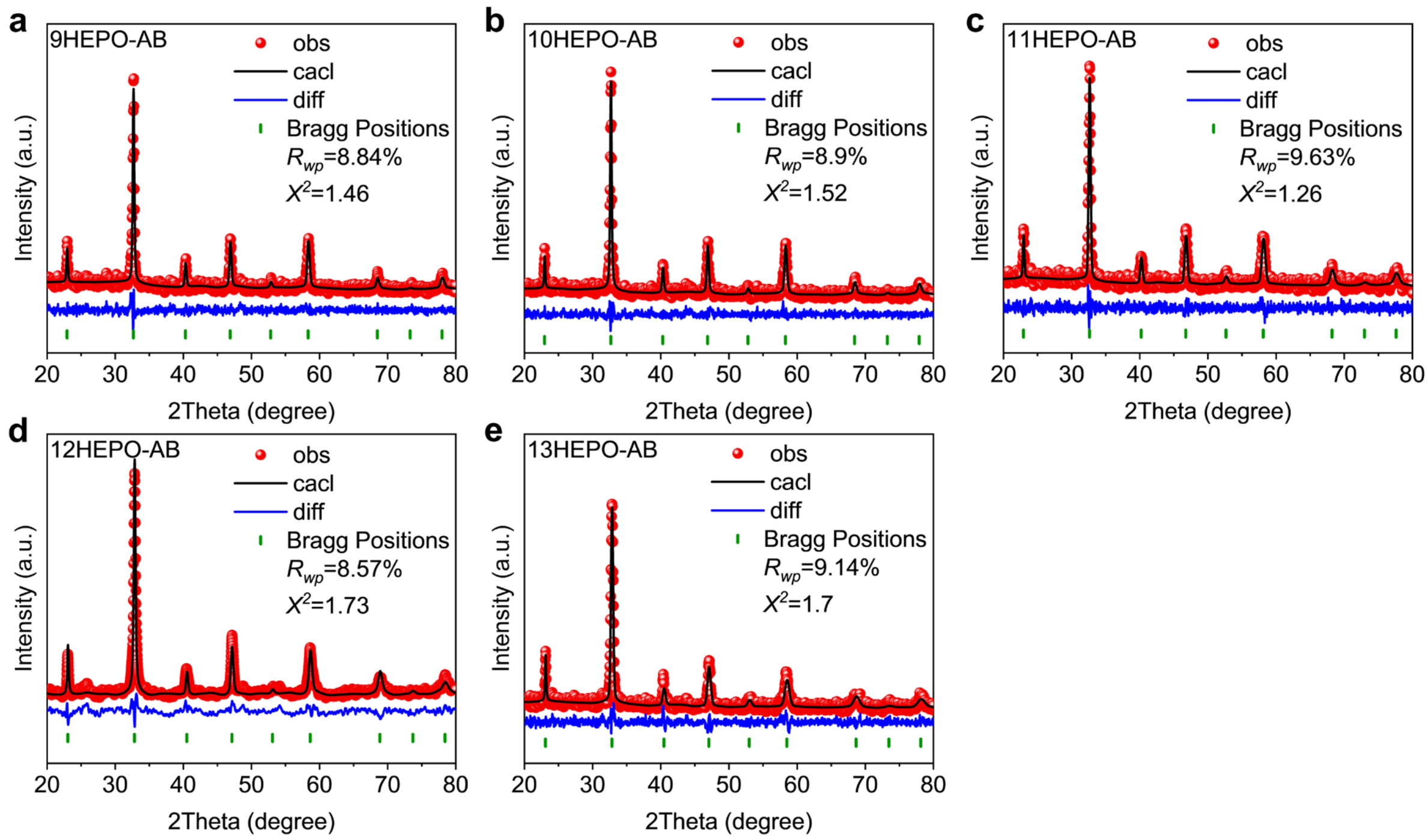

**Figure S15.** XRD Rietveld-refinement profiles of HEPO-AB samples. (a) 9HEPO-AB. (b) 10HEPO-AB. (c) 11HEPO-AB. (d) 12HEPO-AB. (e) 13HEPO-AB.

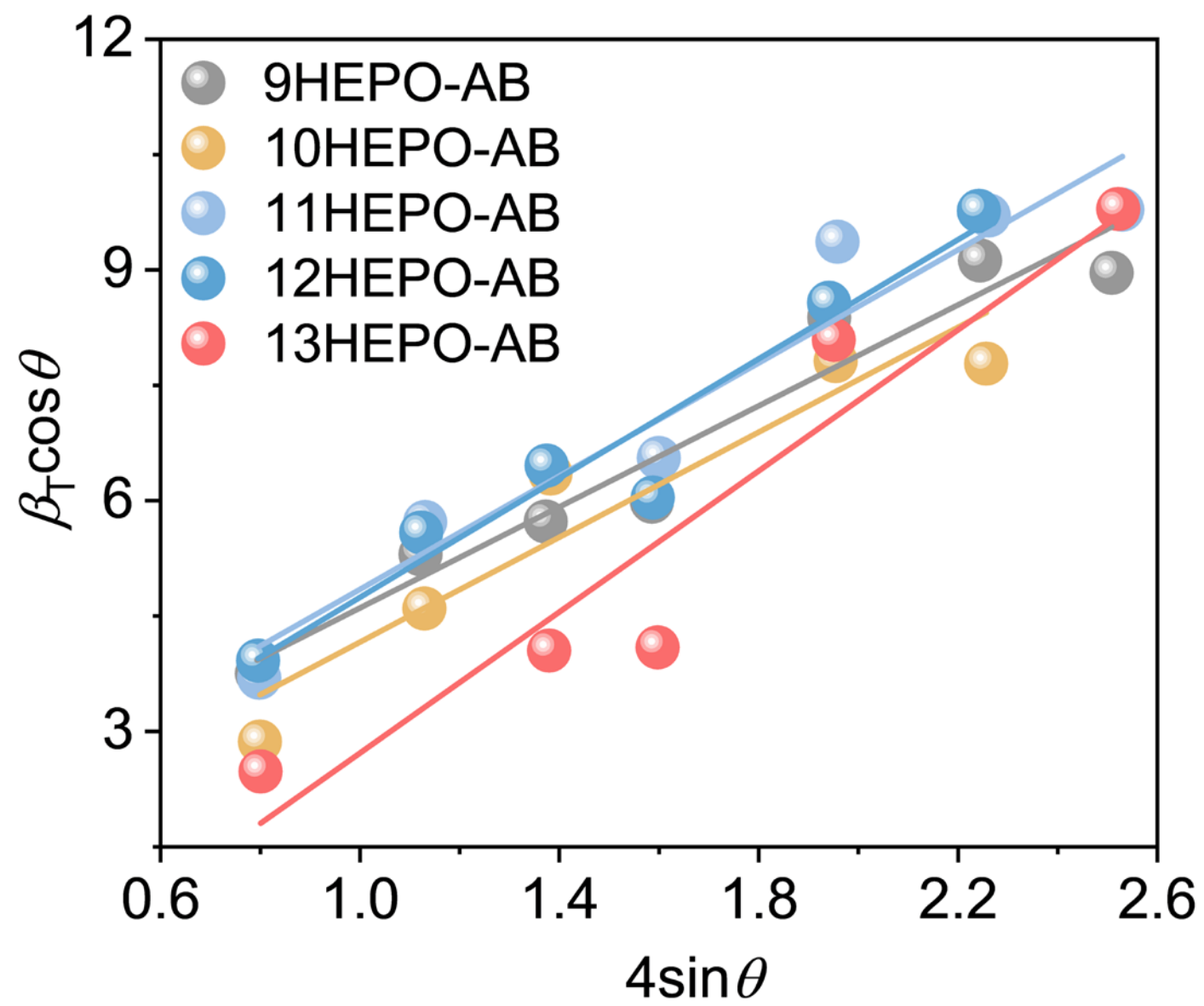

**Figure S16.** $\varepsilon$ of HEPO-AB samples.

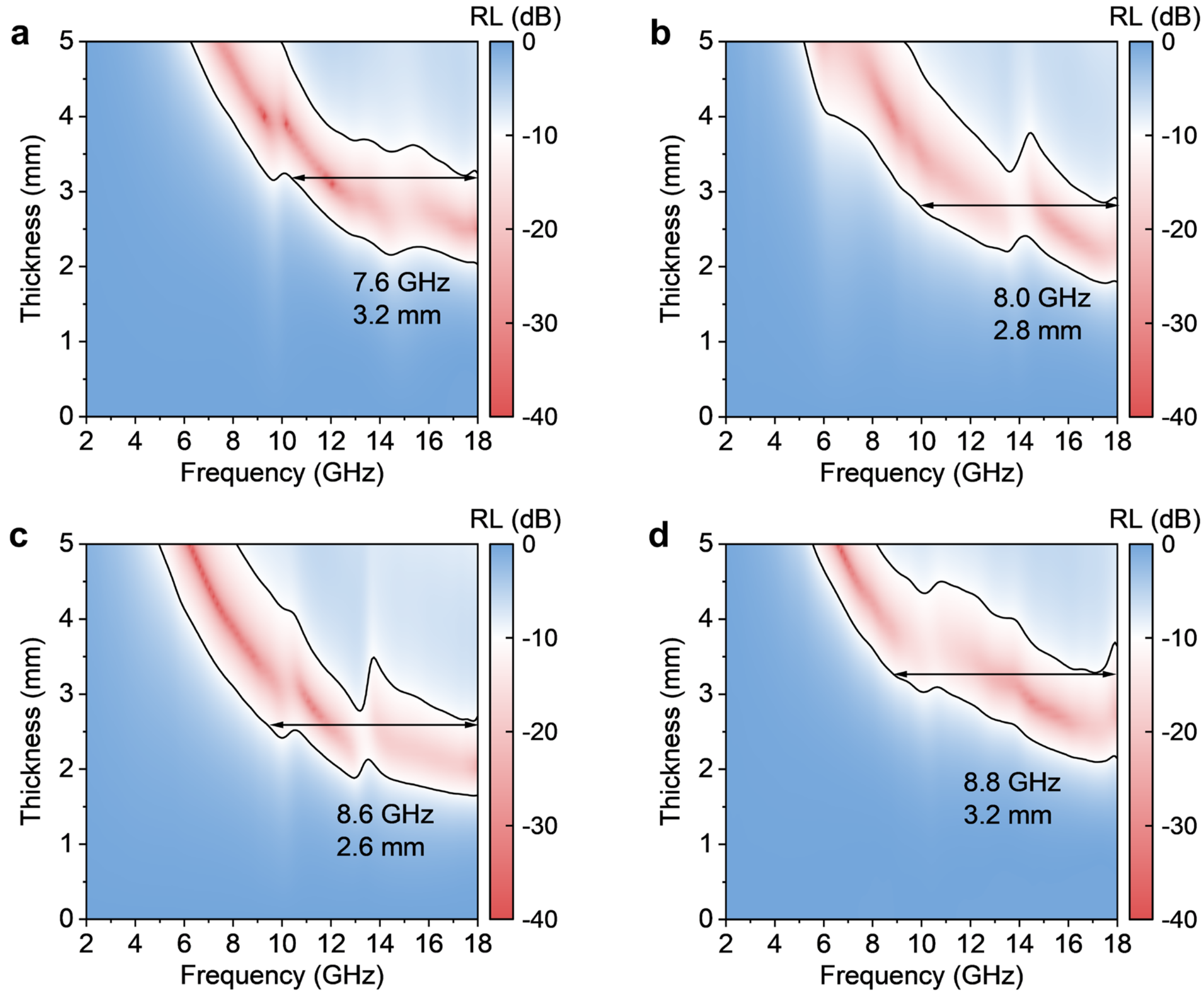

**Figure S17.** EMW absorption performance of HEPO-AB samples. (a) 9HEPO-AB. (b) 10HEPO-AB. (c) 11HEPO-AB. (d) 12HEPO-AB.

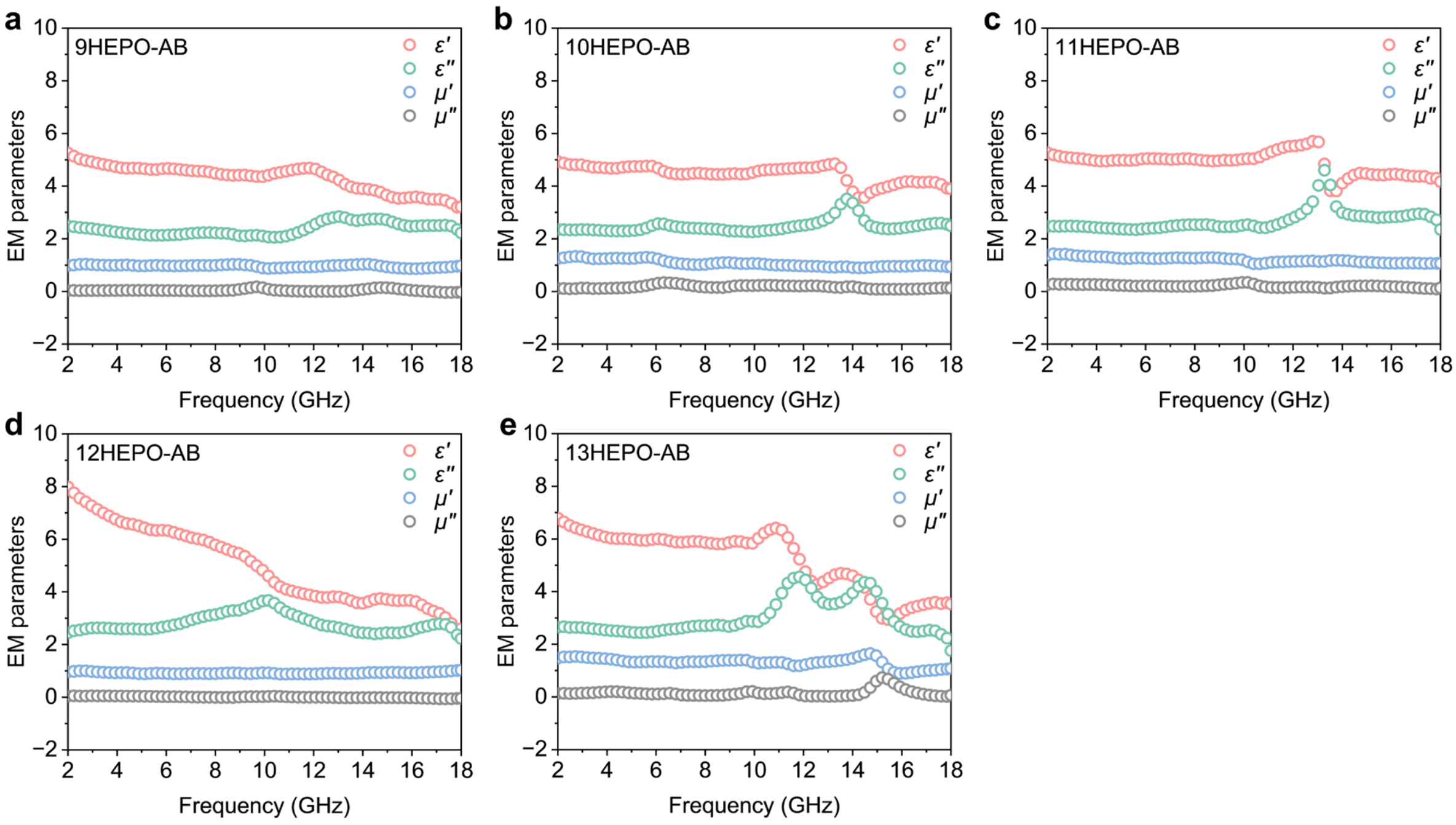

**Figure S18.** EM parameters of HEPO-AB samples. (a) 9HEPO-AB. (b) 10HEPO-AB. (c) 11HEPO-AB. (d) 12HEPO-AB. (e) 13HEPO-AB.

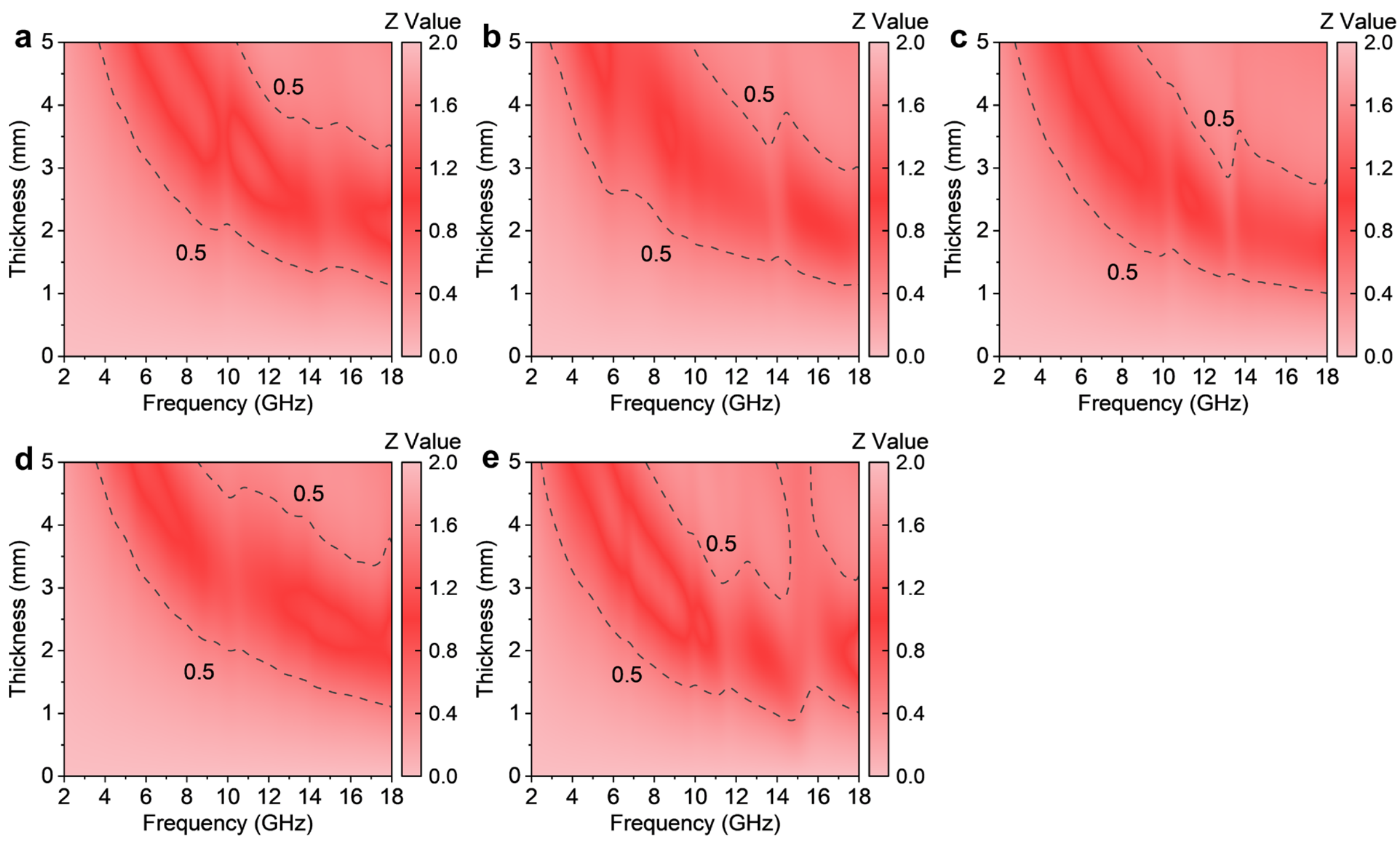

**Figure S19.** $|Z_{in}/Z_0|$ patterns of HEPO-AB samples. (a) 9HEPO-AB. (b) 10HEPO-AB. (c) 11HEPO-AB. (d) 12HEPO-AB. (e) 13HEPO-AB.

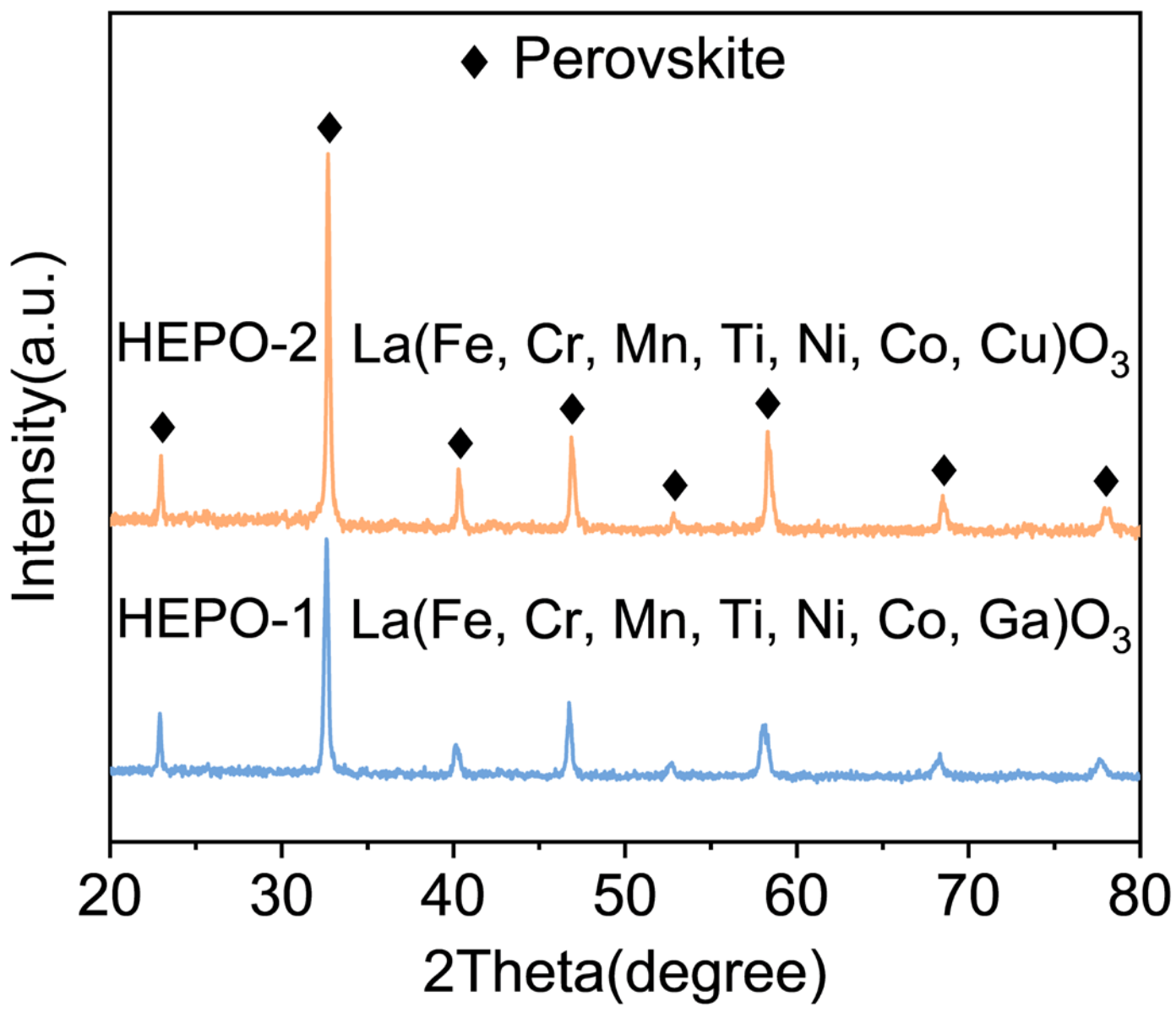

**Figure S20.** XRD patterns of HEPO-1 and HEPO-2.

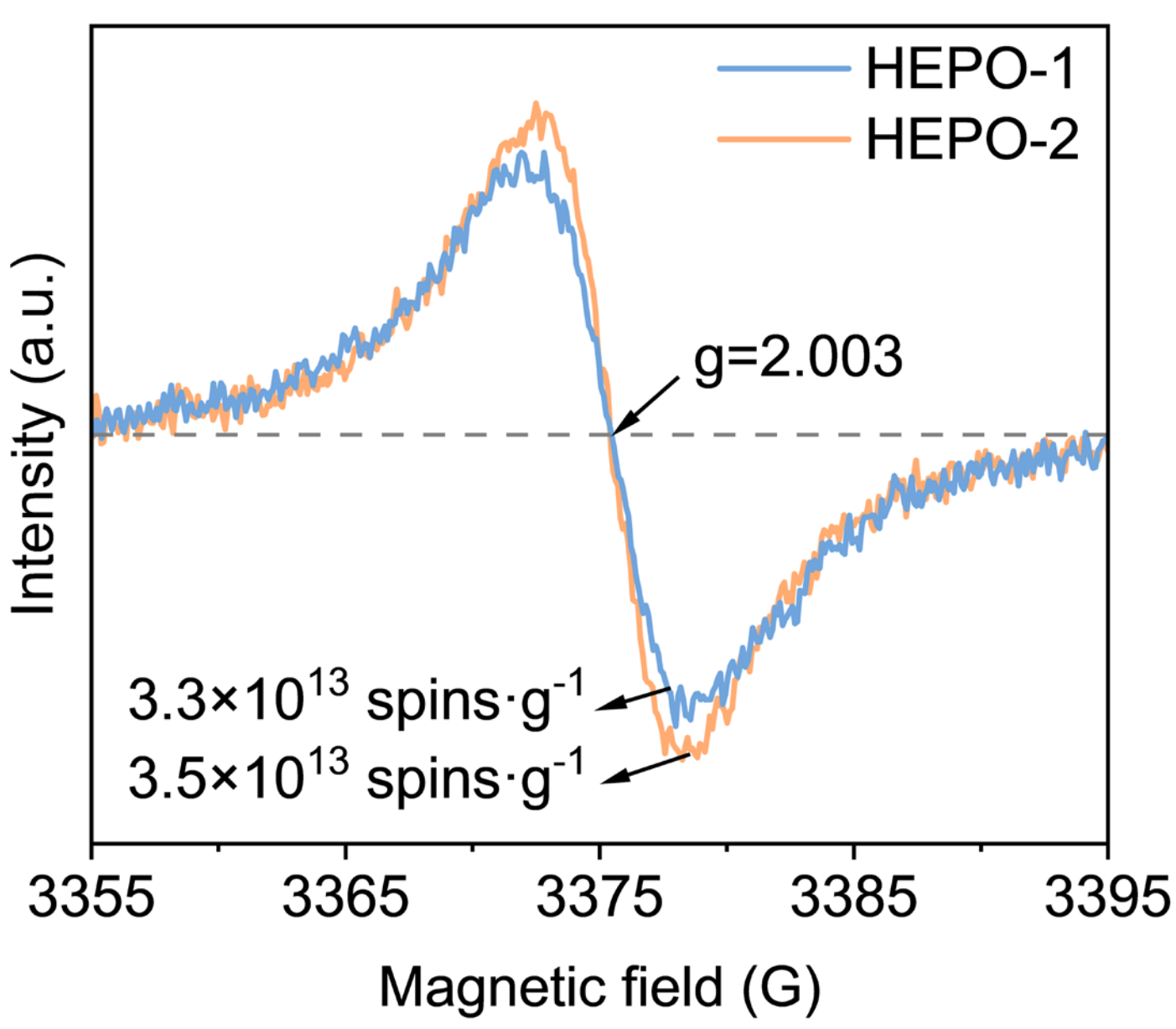

**Figure S21.** Vacancy concentrations of HEPO-1 and HEPO-2.

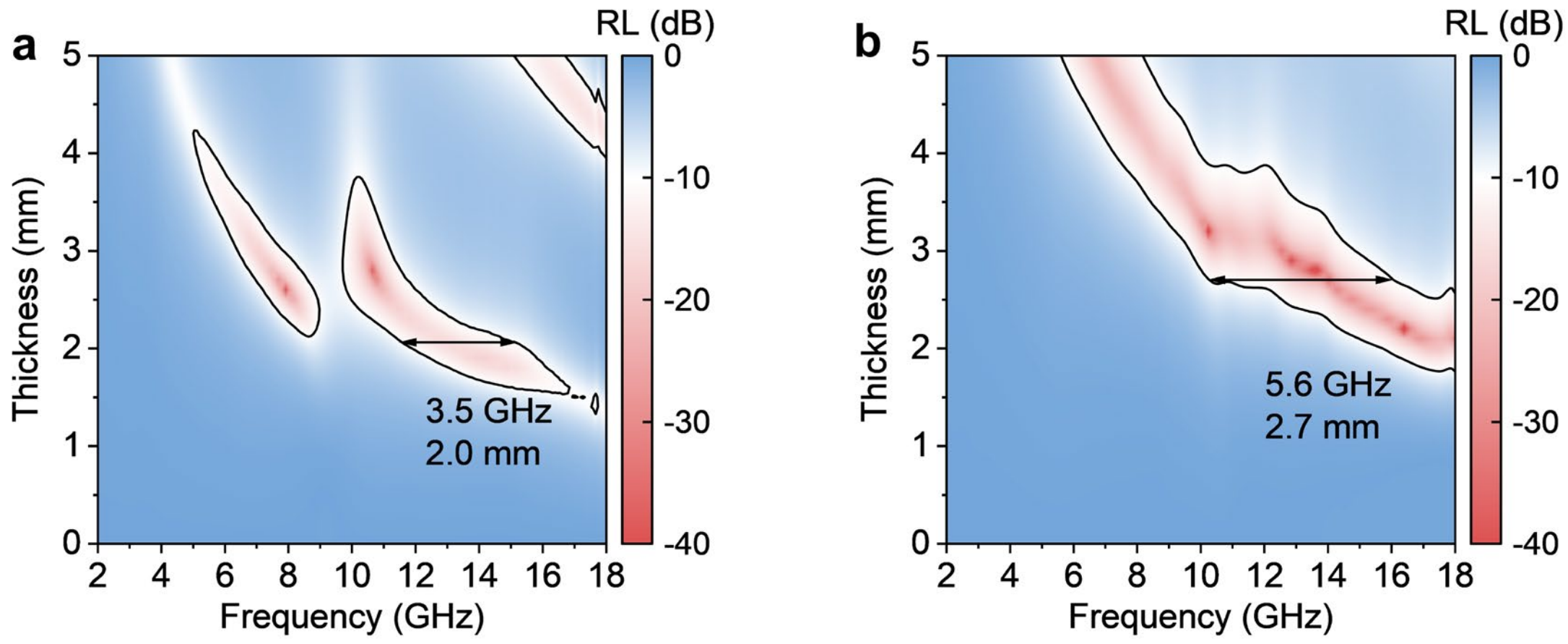

**Figure S22.** EMW absorption performance of (a) HEPO-1 and (b) HEPO-2.

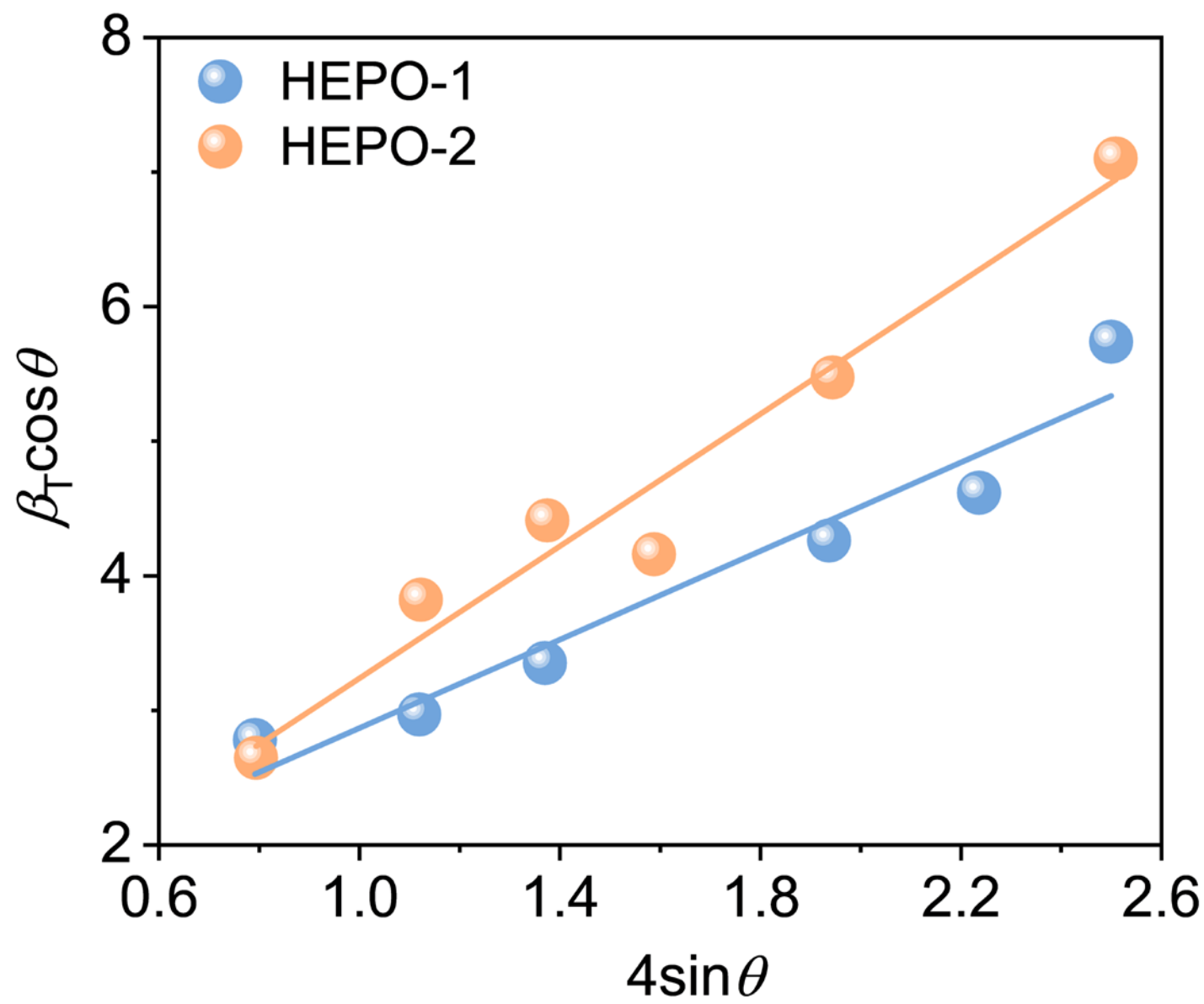

**Figure S23.** $\varepsilon$ of HEPO-1 and HEPO-2.

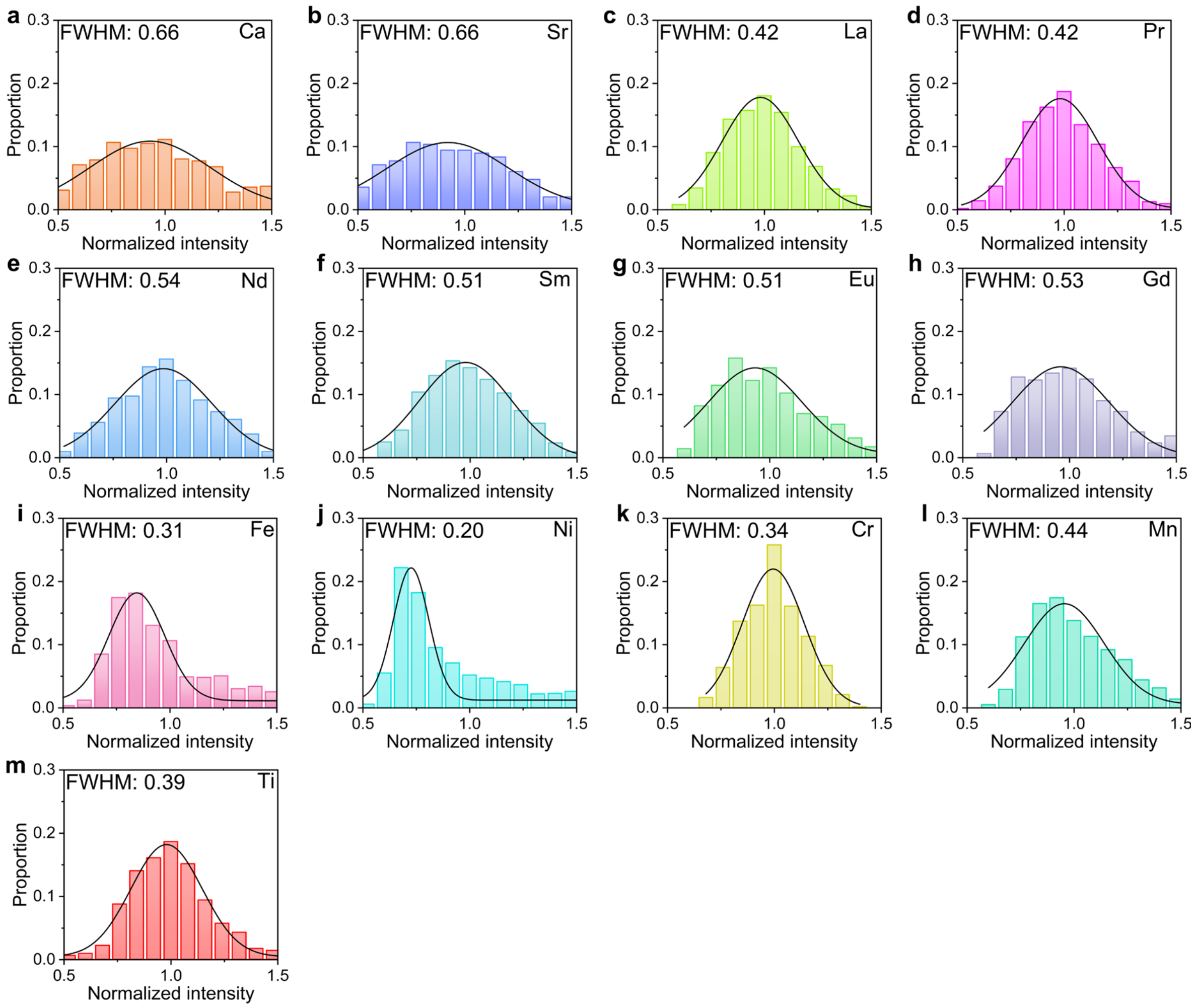

**Figure S24.** Atomic displacements derived from atomic-resolution HAADF-STEM images of 13HEPO-AB. (a) Ca. (b) Sr. (c) La. (d) Pr. (e) Nd. (f) Sm. (g) Eu. (h) Gd. (i) Fe. (j) Ni. (k) Cr. (l) Mn. (m) Ti.

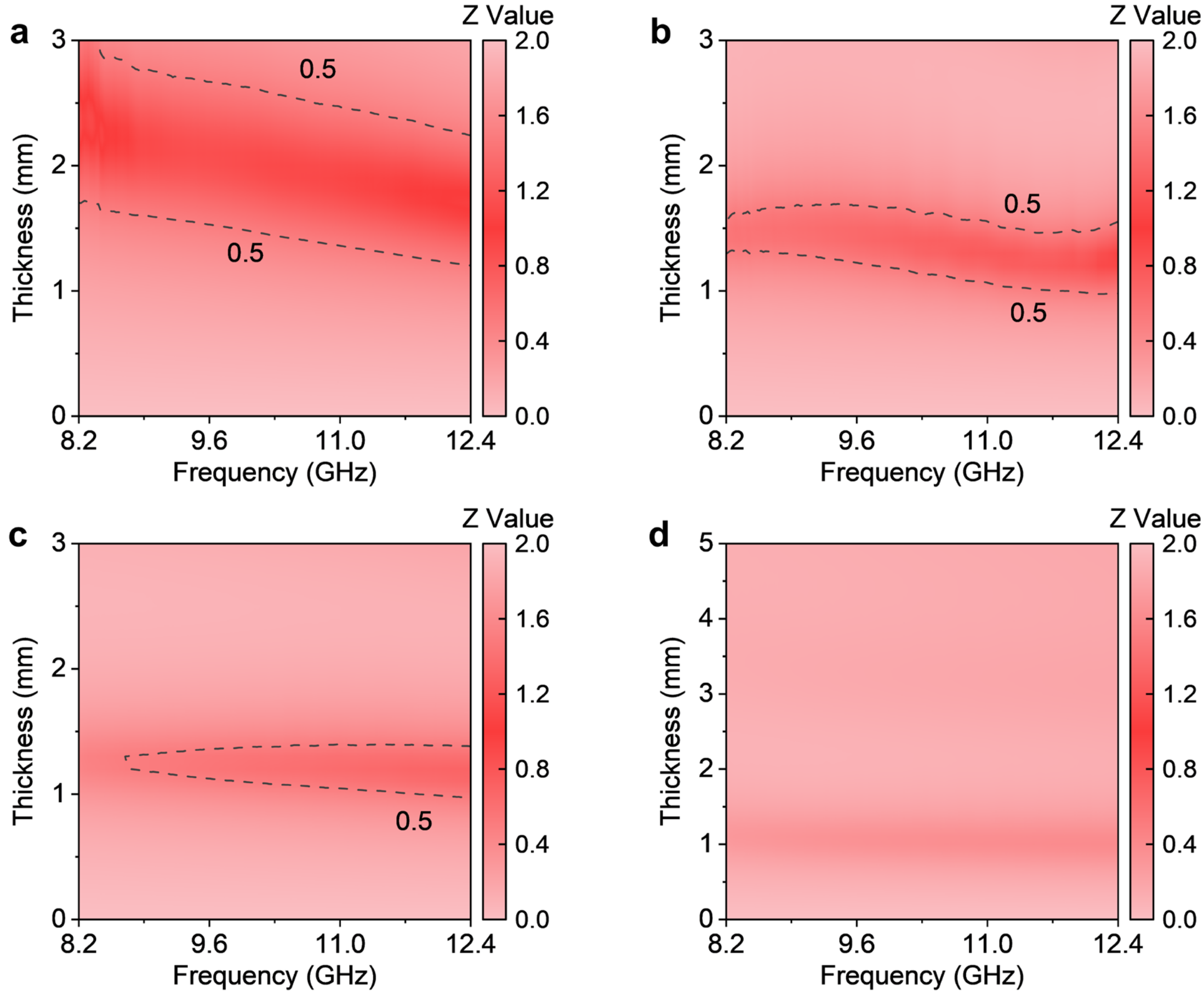

**Figure S25.** $|Z_{in}/Z_0|$ patterns of 13HEPO-AB samples measured across the X-band at different temperatures. (a) RT. (b) 400 °C. (c) 600 °C. (d) 700 °C.

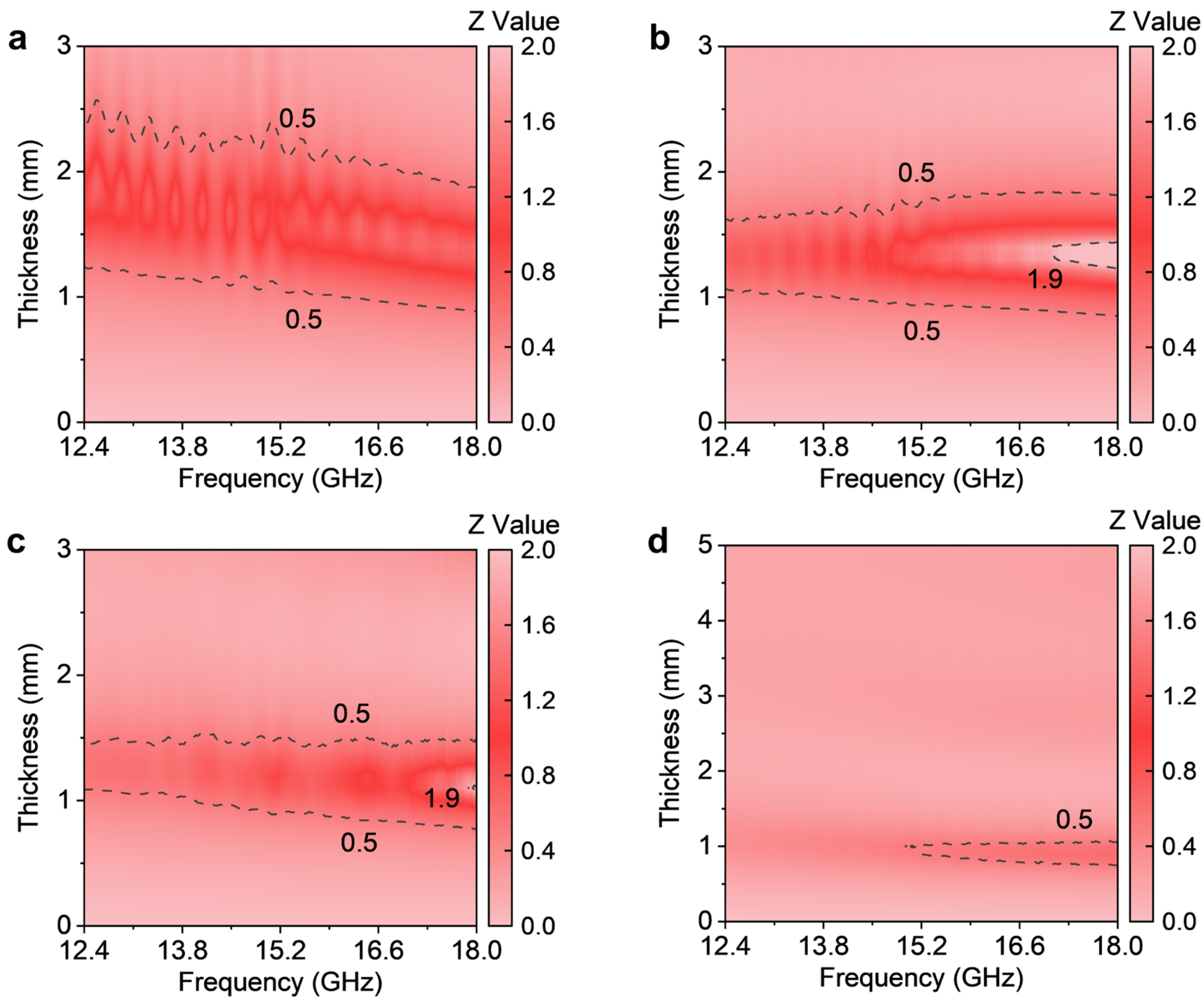

**Figure S26.** $|Z_{in}/Z_0|$ patterns of 13HEPO-AB samples measured across the Ku-band at different temperatures. (a) RT. (b) 400 °C. (c) 600 °C. (d) 700 °C.

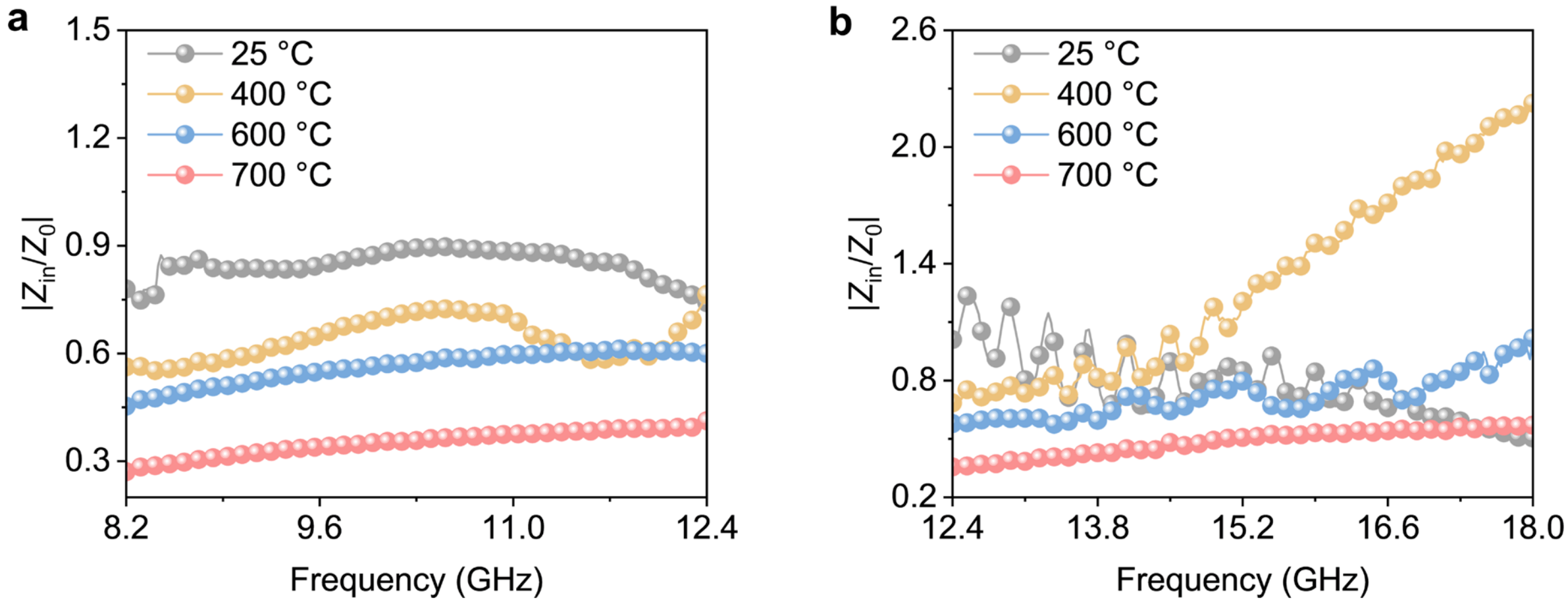

**Figure S27.** $|Z_{in}/Z_0|$ patterns of 13HEPO-AB samples at the optimal thickness measured across different bands. (a) X band. (b) Ku band.

**Table S1.** Nominal formulas of 4–10HEPO-A.

| Number of components | Chemical formula |
|---|---|
| 4 | $(La_{0.25}Sr_{0.25}Ca_{0.25}Ba_{0.25})FeO_3$ |
| 5 | $(La_{0.2}Sr_{0.2}Ca_{0.2}Ba_{0.2}Nd_{0.2})FeO_3$ |
| 6 | $(La_{1/6}Sr_{1/6}Ca_{1/6}Ba_{1/6}Nd_{1/6}Sm_{1/6})FeO_3$ |
| 7 | $(La_{1/7}Sr_{1/7}Ca_{1/7}Ba_{1/7}Nd_{1/7}Sm_{1/7}Pr_{1/7})FeO_3$ |
| 8 | $(La_{1/8}Sr_{1/8}Ca_{1/8}Ba_{1/8}Nd_{1/8}Sm_{1/8}Pr_{1/8}Eu_{1/8})FeO_3$ |
| 9 | $(La_{1/9}Sr_{1/9}Ca_{1/9}Ba_{1/9}Nd_{1/9}Sm_{1/9}Pr_{1/9}Eu_{1/9}Gd_{1/9})FeO_3$ |
| 10 | $(La_{0.1}Sr_{0.1}Ca_{0.1}Ba_{0.1}Nd_{0.1}Sm_{0.1}Pr_{0.1}Eu_{0.1}Gd_{0.1}Er_{0.1})FeO_3$ |

**Table S2.** Nominal formulas of 4–8HEPO-B.

| Number of components | Chemical formula |
|---|---|
| 4 | $La(Fe_{0.25}Cr_{0.25}Mn_{0.25}Ti_{0.25})O_3$ |
| 5 | $La(Fe_{0.2}Cr_{0.2}Mn_{0.2}Ti_{0.2}Ni_{0.2})O_3$ |
| 6 | $La(Fe_{1/6}Cr_{1/6}Mn_{1/6}Ti_{1/6}Ni_{1/6}Co_{1/6})O_3$ |
| 7 | $La(Fe_{1/7}Cr_{1/7}Mn_{1/7}Ti_{1/7}Ni_{1/7}Co_{1/7}Cu_{1/7})O_3$ |
| 8 | $La(Fe_{1/8}Cr_{1/8}Mn_{1/8}Ti_{1/8}Ni_{1/8}Co_{1/8}Cu_{1/8}Al_{1/8})O_3$ |

**Table S3.** Nominal compositions of 9–13HEPO-AB.

| Number of components | Chemical formula |
|---|---|
| 9 | $(La_{0.2}Sr_{0.2}Ca_{0.2}Nd_{0.2}Sm_{0.2})(Fe_{0.25}Cr_{0.25}Mn_{0.25}Ti_{0.25})O_3$ |
| 10 | $(La_{0.2}Sr_{0.2}Ca_{0.2}Nd_{0.2}Sm_{0.2})(Fe_{0.2}Cr_{0.2}Mn_{0.2}Ti_{0.2}Ni_{0.2})O_3$ |
| 11 | $(La_{1/6}Sr_{1/6}Ca_{1/6}Nd_{1/6}Sm_{1/6}Pr_{1/6})(Fe_{0.2}Cr_{0.2}Mn_{0.2}Ti_{0.2}Ni_{0.2})O_3$ |
| 12 | $(La_{1/7}Sr_{1/7}Ca_{1/7}Nd_{1/7}Sm_{1/7}Pr_{1/7}Eu_{1/7})(Fe_{0.2}Cr_{0.2}Mn_{0.2}Ti_{0.2}Ni_{0.2})O_3$ |
| 13 | $(La_{0.125}Sr_{0.125}Ca_{0.125}Nd_{0.125}Sm_{0.125}Pr_{0.125}Eu_{0.125}Gd_{0.125})(Fe_{0.2}Cr_{0.2}Mn_{0.2}Ti_{0.2}Ni_{0.2})O_3$ |

**Table S4.** Benchmarking the key EMW absorption properties of optimal 13HEPO-AB against other reported absorbers at RT.

| Material | EAB (GHz) | $RL_{min}$ (dB) | Thickness (mm) | Reference |
|---|---|---|---|---|
| NiO | 5.6 | −37.1 | 2.5 | [1] |
| $Fe_3O_4$ | 4.8 | −40.4 | 5.9 | [2] |
| $(Mg_{0.2}Co_{0.2}Ni_{0.2}Cu_{0.2}Zn_{0.2})O$ | 2.92 | −41.4 | 2 | [3] |
| $(Fe_{0.2}Co_{0.2}Ni_{0.2}Cr_{0.2}Mn_{0.2})_3O_4$ | 6.2 | −54.1 | 1.5 | [4] |
| $(MnNiCuZn)_{0.7}Co_{0.3}Fe_2O_4$ | 2.5 | −27 | 5.3 | [5] |
| $La_3Ni_2O_7/LaNiO_3$ | 3.36 | −43.3 | 2.2 | [6] |
| $(Fe_{0.2}Co_{0.2}Ni_{0.2}Cu_{0.2}Zn_{0.2})O/$ $(Fe_{0.2}Co_{0.2}Ni_{0.2}Cu_{0.2}Zn_{0.2})Fe_2O_4$ | 3.3 | −54.5 | 2 | [7] |
| $BaFe_{11.6}Gd_{0.4}O_{19}$ | 7.83 | −33.9 | 2 | [8] |
| $BaLa_{0.1}Ce_{0.1}Fe_{11.8}O_{19}$ | 4.04 | −56.2 | 1.5 | [9] |
| $La(Fe_{1/3}Co_{1/3}Ni/_{1/3})O_3$ | 3.2 | −45.8 | 2.2 | [10] |
| $Sr(Cr_{0.2}Mn_{0.2}Fe_{0.2}Co_{0.2}Ni_{0.2})O_3$ | 7.44 | −54 | 1.8 | [11] |
| $Pr_{0.98}Ba_{0.02}MnO_3$ | 6.16 | −46.4 | 3.7 | [12] |
| $(Ba_{1/3}Ca_{1/3}Sr_{1/3})FeO_3$ | 4.16 | −40.5 | 1.6 | [13] |
| $La(Fe_{0.2}Co_{0.2}Ni_{0.2}Cr_{0.2}Mn_{0.2})O_3$ | 7.6 | −50.6 | 2.78 | [14] |
| 13HEPO-AB | 9.2 | −50 | 2.7 | This work |

**Table S5.** Comparison of microwave-absorbing performance of high-temperature microwave-absorbing materials reported in recent years.

| Material | EAB (GHz) | $RL_{min}$ (dB) | Thickness (mm) | Temperature (°C) | Reference |
|---|---|---|---|---|---|
| $(FeCoNiCrMn)_3O_4$ | 2 | −22.5 | 1.9 | 600 | [4] |
| $La_{0.7}Sr_{0.3}Mn_{0.8}Fe_{0.2}O_3$/ MAS | 4.2 | −17.9 | 1.9 | 500 | [15] |
| $Er_2Zr_2O_7/Gd_2Zr_2O_7$ | 8.27 | −35.1 | 1.1 | 600 | [16] |
| $Ti_3SiC_2/Al_2O_3$-13%$TiO_2$ | 2.12 | −51.8 | 2.2 | 500 | [17] |
| $ZnO/SiC_{nw}$ | 3.6 | −40.2 | 3.6 | 600 | [18] |
| $SiC_f/SiC$-$Al_2O_3$ | 4.1 | −22.8 | 3 | 400 | [19] |
| $CNT_s/ZnO$ | 3.2 | −22 | 2.7 | 400 | [20] |
| $Ti_3AlC_2/La_2Zr_2O_7$ | 2.4 | −56.8 | 2.7 | 600 | [21] |
| $Si_3N_4$-$SiC/SiO_2$ | 4.04 | −18.0 | 3.3 | 600 | [22] |
| $ZnO/CNT/SiO_2$ | 3.4 | −13.0 | 2.5 | 400 | [23] |
| NiO/SiC | 4.2 | −46.9 | 2 | 500 | [24] |
| $MgO/TiB_2$ | 4.2 | −52.1 | 1.6 | 500 | [25] |
| $CNT_s/Sc_2Si_2O_7$ | 4.2 | −47.5 | 2.8 | 400 | [26] |
| $TiN/BN/SiO_2$ | 2.71 | −16.7 | 3 | 600 | [27] |
| Graphenen/$Fe_3O_4$/ SiBCN | 3.93 | −45.6 | 2.1 | 600 | [28] |
| $Co_3O_4/rGO/SiO_2$ | 4.2 | −29.5 | 1.7 | 200 | [29] |
| 13HEPO-AB | 9.7 | −25.8 | 4 | 600 | This work |

**Supplementary references**